\documentclass[prodmode,acmtochi]{acmsmall} 

\usepackage{balance}
\usepackage{graphicx}
\usepackage{epstopdf}
\usepackage{times}
\usepackage{url}
\usepackage{cite}
\usepackage{float}
\usepackage{subfig}
\usepackage{tabularx}
\usepackage{enumerate}
\usepackage{tikz}
\usepackage{amsmath}
\usepackage{booktabs}
\usepackage{colortbl}
\usepackage{pdfpages}
\usepackage{longtable}
\usepackage[ruled]{algorithm2e}
\renewcommand{\algorithmcfname}{ALGORITHM}
\SetAlFnt{\small}
\SetAlCapFnt{\small}
\SetAlCapNameFnt{\small}
\SetAlCapHSkip{0pt}
\IncMargin{-\parindent}

\graphicspath{{./Figures/}}

\makeatletter

\newcommand{\Rmnum}[1]{\expandafter\@slowromancap\romannumeral #1@}
\makeatother

\begin{document}

\markboth{A. Jahanian et al.}{Colors ---Messengers of Concepts: Visual Design Mining for Learning Color Semantics}

\title{Colors ---Messengers of Concepts: Visual Design Mining for Learning Color Semantics}

\author{Ali Jahanian
\affil{Purdue University}
S.V.N. Vishwanathan
\affil{Purdue University}
Jan P. Allebach
\affil{Purdue University}
}

\begin{abstract}
This paper studies the concept of color semantics by modeling a dataset of magazine cover designs, evaluating the model via crowdsourcing, and demonstrating several prototypes that facilitate color-related design tasks. We investigate a probabilistic generative modeling framework that expresses semantic concepts as a combination of color and word distributions ---color-word topics. We adopt an extension to Latent Dirichlet Allocation (LDA) topic modeling called LDA-dual to infer a set of color-word topics over a corpus of 2,654 magazine covers spanning 71 distinct titles and 12 genres. While LDA models text documents as distributions over word topics, we model magazine covers as distributions over color-word topics. The results of our crowdsourced experiments confirm that the model is able to successfully discover the associations between colors and linguistic concepts. Finally, we demonstrate several simple prototypes that apply the learned model to color palette recommendation, design example retrieval, image retrieval, image color selection, and image recoloring.

\end{abstract}




\acmformat{Ali Jahanian, S.V.N. Vishwanathan, and Jan P. Allebach, 2014. Colors ---Messengers of Concepts: Visual Design Mining for Learning Color Semantics}

\begin{bottomstuff}

Author's addresses: A. Jahanian (current address) MIT, jahanian@csail.mit.edu; S.V.N. Vishwanathan, (current address) Jack Baskin School of Engineering, Computer Science Department, University of California Santa Cruz, vishy@ucsc.edu; J. P. Allebach, School of Electrical and Computer Engineering, Purdue University, allebach@purdue.edu.
\end{bottomstuff}

\maketitle

\section{Introduction}
\label{sec:Introduction}

\begin{figure}[h!tb]
  \centering
  \includegraphics[width=.9\textwidth]{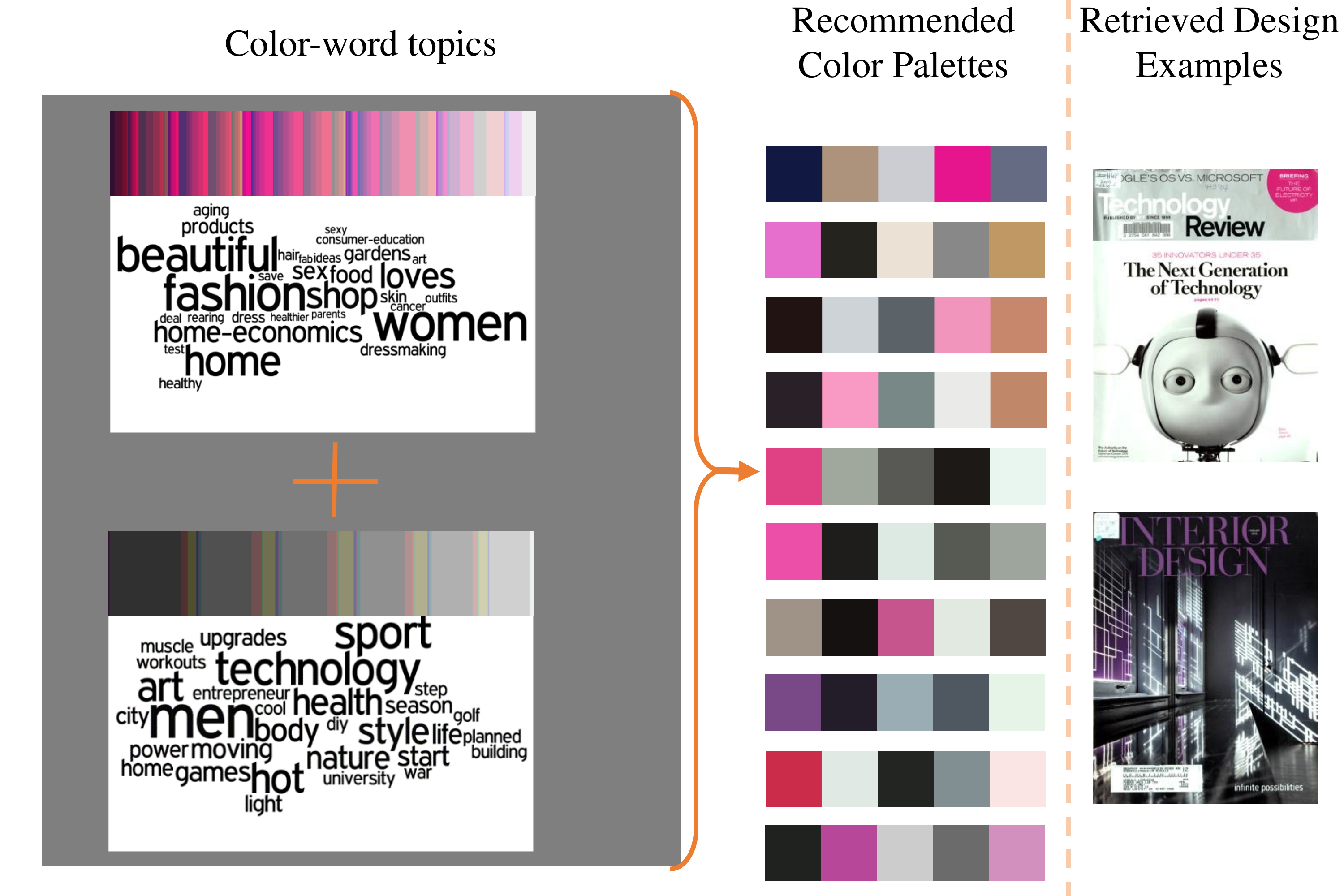}
  \caption{Application of color semantics in color palette selection, and design example retrieval. See Sec.~\ref{sec:Applications}.}
  \label{teaser_img}
\end{figure}

Although researchers have endeavored to understand the role of color perception in design of human computer interactions, in user engagement and first impression~\cite{lindgaard2006attention,reinecke2013predicting}, trust and credibility~\cite{lindgaard2011exploration,pengnate2013multimethod}, performance~\cite{moshagen2009blessing,sonderegger2010influence}, aesthetics and usability~\cite{lavie2004assessing,hassenzahl2004interplay}, and culture~\cite{cyr2010colour,reinecke2011improving,reinecke2014}, understanding the cognitive aspect of color merits further investigation.
It is more evident in the recent studies~\cite{derefeldt2004cognitive,skorupski2011colour} that colors are cognitive: Perceived by
our visual system, colors are classified at higher levels of
abstraction into verbal and semantic
categories~\cite{humphreys1989visual,barsalou1999perceptual} by a visual
task such as color categorization~\cite{derefeldt2004cognitive}. Nevertheless, in practice, for instance in visual design, color combinations are designed not only to be
appealing, but also to be silent salespersons that communicate with the
audience~\cite{eisemann2000pantone,frascara2004communication,newark2007graphic,samara2007design}.
In schools of design, students learn creation and usage of color
palettes to convey emotions and
ideas~\cite{whitfield1982design,ou2012cross}.
Devising generative models which can associate color combinations with linguistic concepts, based on the work of designers, might enable more meaningful user interactions for communicating with colors through color design and cognition.

Color semantics in computer science and engineering has been observed as a
gap~\cite{smeulders2000content,sethi2001mining,mojsilovic2001capturing,liu2007survey}.
Nevertheless, in terms of associating emotions and moods with colors, there is a recent body of work~\cite{csurka2010learning,solli2010color,murray2012toward}. In current frameworks, the semantics are labels provided by human inputs, leading to a manual collection and a limited number of semantics. However, the task of automatically inferring the linguistic concepts from the data is still unsolved. Another limitation is that these semantics are defined in bipolar scales, such as \emph{happy-sad}. This limits the notion of color semantics to the opposites. In other words, more abstract concepts, such as \emph{management} will not be captured.

In this paper, we investigate a probabilistic generative modeling framework to associate color combinations with linguistic concepts from professional design examples. The model is an adaptation of LDA-dual, an extension of the popular Latent Dirichlet Allocation (LDA) topic model~\cite{blei2003latent} to jointly link colors and words, and infer \emph{color-word topics}.
Originally, the LDA-dual model was proposed by Shu et al.~\cite{shu2009latent} for complete entity resolution of names in the bibliography of scientific publications.
Because LDA is both an inference mechanism and a generative model, our framework utilizes both these features to automatically infer semantics, and to enables more meaningful user interactions for color selection and design. As one application, Fig.~\ref{teaser_img} shows a system for finding design examples based on color semantics.

In visual design, designers use both colors and words as design elements to carry their messages in a unified form of design.
This observation suggests that each design's
theme might be a combination of words and color distributions; and each
design may include a proportion of various themes.  A similar intuition
is argued in LDA~\cite{blei2003latent}, for textual documents.  That is, a document can be
viewed as a proportion of different topics, where each topic is a histogram of words.
Although one may adopt LDA to capture the underlying word topics or color topics in design examples, our aim is to simultaneously link both of these topics via our framework.
The model produces combined color-word topics. The word topics are multinomial distributions on words; and the color topics are histograms of colors.
Later, to facilitate the user interactions, we visualize the word topics as word clouds, and discretize the resulting color histograms to yield a mapping to 5-color palettes.

Seeking good examples of visual design, we collected a relatively large
corpus of magazine covers.  We chose magazine covers for two reasons:
First, by using a variety of data sources including the Internet,
archives of our own university library, and widely accessible
newsstands, we were able to create a dataset of 2,654 covers from 71
titles and 12 genres.  Second, designers spend many days conceptualizing and creating
a magazine cover that attracts the audience at first glance, while
competing with other magazines on a newsstand~\cite{magbook}.  In this
process, designers carefully choose a color palette for the cover based on
the topic and cover lines of the issue in order to convey a very specific concept.

To verify  whether or not users agree with the associations between color combinations and linguistic concepts produced by the model, we conduct a crowdsourced experiment. In particular, users are shown discretized color palettes and a choice of word clouds and are asked to select the most appropriate word cloud for the given color palette.  Based on the user feedback, we infer the relevance of the color palette to the word cloud in the experiment.

The flow of this paper is as follows. In Sec.~\ref{sec:Theory}, we present a study on color semantics in both theoretical and practical aspects. This section is then followed by Sec.~\ref{sec:RelatedWork} where we review related work in computer science and engineering. In Sec.~\ref{sec:DataCollection}, we discuss how we preprocess the data, and define vocabularies for words and the basis that we use to represent colors in our system. We then discuss the inference and generative mechanisms in the modeling framework in Sec.~\ref{sec:StatisticalModel}. For ease of visualization, in Sec.~\ref{sec:InterpretingModelOutput} we represent the word topics as word clouds, and discretize the resulting color histograms to yield a mapping to a pool of 5-color palettes. We then discuss the design of the crowdsourcing experiments in Sec.~\ref{sec:UserStudy}, and analyze the crowd responses in Sec.~\ref{sec:InterpretingTheUserStudy}. In Sec.~\ref{sec:Applications}, we suggest a number of applications for color semantics to demonstrate how semantics can enable more meaningful user interactions.
We specifically discuss color palette selection and design example recommendation, image retrieval, color region selection in images, and pattern colonization in image recoloring. We conclude this paper and suggest a number of interesting avenues for future work in Sec.~\ref{sec:ConclusionAndFutureWork}.

\section{Theory of Color Semantics}
\label{sec:Theory}

Colors are colorful: more than physical attributes of objects, colors are cognitive and have names, meanings, and values. In the following subsections, these aspects of colors are briefly reviewed, and the need to understand the role of color semantics in analysis and synthesis of visual designs in HCI is elucidated.

\subsection{Color Cognition}

While color perception is a rich area of research, color cognition is relatively new~\cite{pylyshyn1999vision}.
That is, although color is perceived as the result of wavelength discrimination, it is also cognitive~\cite{skorupski2011colour}.
Derefeldt et al.~\cite{derefeldt2004cognitive} elucidate that ``cognitive'' means that after visual perception, color is classified to a higher level of abstraction into verbal and semantic categories~\cite{humphreys1989visual,barsalou1999perceptual} by a visual task such as color categorization.

Heer and Stone~\cite{heer2012color} suggest that when this categorization capability (in the concept of color naming) is deployed in user interfaces that model human category judgements, it might demonstrate more meaningful and novel user interactions. What the boundaries are for color naming, and how this concept can be compared with color meanings has not yet received the attention of scientists.

\subsection{Color Naming}

The earliest effort to understand linguistic categorization of colors is perhaps the work of Berlin and Kay for elucidating that color naming is universal and related to evolution~\cite{berlin1991basic}. Before their work, most linguists believed that the concept of color naming ---associating names with colors--- is not universal. As Kay and McDaniel~\cite{kay1978linguistic} summarize, the two main misconceptions about color naming among linguists were that color naming is a matter of cultural relativism and that semantic primes in languages are discrete entities. The latter doctrine was not able to address compound terms, e.g. \emph{green-blue}. This inadequacy motivated Kay and McDaniel to describe color categories as continuous functions using fuzzy set theory.
The universality claim of color naming~\cite{berlin1991basic} does pertain under some conditions, as explained by Palmer in his book \emph{Vision Science: Photons to Phenomenology}~\cite{palmer1999vision}. Although color naming associates words with colors, there are other trends of research that investigate meanings and higher level abstractions/semantics of colors.

\subsection{Color Meanings and Semantics}

The first systematic approach to quantifying ``meanings'' of linguistic concepts comes from \emph{measurement of meaning} by Osgood~\cite{osgood1952nature,osgood1957measurement}. Osgood proposes a semantic space based on pairs of polar terms, such as \emph{happy-sad} or \emph{kind-cruel}, and terms it a ``semantic differential'' as an ``objective index of meaning''~\cite{osgood1952nature}. Osgood then studies cross-cultural generality of his semantic space and finds significant similarities between different primitive cultures~\cite{osgood1960cross}. Later, Osgood compiled a 620-concept \emph{Atlas of Affective Meanings}, explained in~\cite{osgood1971exploration} with a report of cross-cultural studies in 23 cultures. Adams and Osgood~\cite{adams1973cross} investigate the 8 color concepts of Osgood's Atlas ---\emph{color} (vs monochrome), \emph{white}, \emph{grey}, \emph{black}, \emph{red}, \emph{yellow}, \emph{green}, and \emph{blue}--- among 20 countries and report that for instance, the relative affective meaning of \emph{red} is \emph{strong} and \emph{active}.
Another similar attempt is due to Wright and Rainwater~\cite{wright1962meanings}, yet with a more visual communication language perspective in color meanings and connotations. This is where a set of more design-oriented color semantics such as ``elegant'' or ``showiness'' emerges.

A breakthrough in design-based color semantics is the study by Kobayashi~\cite{kobayashi1981aim}. Kobayashi defines a meaning scale to relate ``worlds of people and objects with worlds of colors'' and terms it \emph{Color Image Scale}~\cite{kobayashi1991color}. This scale is comprised of two bipolar dimensions: \emph{warm-cool} and \emph{soft-hard}. Kobayashi's scale has significant contributions to visual design: It is based on 3-color combinations (3-color palettes) rather than single color patches (swatches). It also relates these 3-color combinations with two levels of abstractions, one with 180 semantics, and the other with 15 higher level semantics. Because of these characteristics, Kobayashi's scale is taught in many color courses in schools of design (e.g. see~\cite{green2006value}). Later cross-cultural studies acknowledge the universality of Kobayashi's bipolar scales~\cite{ou2004studyI,ou2004studyII,ou2012cross}.
A similar effort in association of colors and words was undertaken by Sivik as mentioned in~\cite{hardin1997color}.

\subsection{Color Semantics, Emotions, and Preferences}

While investigating meanings of colors, some researchers have attempted to understand the emotions and moods evoked by colors and how feelings about colors can influence our performance (e.g. see~\cite{thuring2007usability} and~\cite{linSemantically2013}). From a psychological point of view for instance, Crozier~\cite{crozier1996psychology} argues that theories of preference based on innate and learned reactions should be considered while studying color meanings. Crozier summarizes that \emph{red}, for instance, has the innate mood of the alert signal and also is involved in sexual behaviour in many species; or \emph{white} is learned to be associated with purity in some cultures. Although Crozier acknowledges that the \emph{like-dislike} bipolar scale in prior studies is a useful measure to investigate color moods, he argues that meanings of individual colors should be considered within the context (syntax, semantics, and culture) in which they are examined, also adding other factors such as age and gender.

Later, Ou et al.~\cite{ou2004studyI,ou2004studyII,ou2004studyIII} studied color emotions and color preferences to clarify the relations between them. Although their studies agreed with prior work in defined scales of \emph{warm/cool}, \emph{heavy/light}, and \emph{active/passive}, they found notable differences in the \emph{like-dislike} scale between their two groups of participants, who were Chinese and British. They also found a tendency for their participants to prefer color combinations that hold opposite emotions.

\subsection{Color Semantics and Cross-cultural Considerations}

Hutchings summarizes that ``The Principle of Adaptation of Ideas accounts for regional variations in colour folklore. This embodies a Darwinian-type principle of behavior, that is, `to survive within a community, a belief must have relevance to that community'''~\cite{hutchings2004colour}. Nevertheless, it is observed that in the information era, common sense about colors is increasing~\cite{carroll2007art}.
Such a cultural problem in HCI can be thought of as ``minor science'' as~\cite{ghassan2013legitimacy} suggest.

In a recent cross-cultural study of color emotions of 190 two-color combinations from inhabitants of eight countries: \emph{Argentina, France, Germany, Iran, Spain, Sweden, Taiwan,} and \emph{UK}, Ou et al.~\cite{ou2012cross} report consistency for \emph{warm/cool}, \emph{heavy/light}, and \emph{active/passive}. This study also points out some inconsistency for \emph{like/dislike}. For instance, Argentinian participants preferred more grayish colors in contrast to other participants. However, additional cross-cultural studies could shed light on our knowledge in color semantics and personal preferences.

\subsection{Color Semantics in Applied Arts}

Krippendorff states that ``Design is making sense (of things)''~\cite{krippendorff1989essential}. This sense making in applied arts is often equated with communication via elements of design. Professional designers and researchers have specifically emphasized the role of colors and color semantics as a means of visual language~\cite{green2006value,hutchings1995continuity}. In architecture~\cite{hogg1979dimensions,caivano2006research,ural2010architectural}, interior design~\cite{poldma2009learning}, textile design~\cite{lee2006development}, and product design in marketing~\cite{madden2000managing}, understanding values of colors and color semantics is acknowledged as a knowledge that can support not only non-designers, but also professional designers who conceptualize and ideate designs based on their own intuitions. This knowledge can enhance the processes involved in design, such as inspiring, brainstorming, exemplifying, and communicating among designers over design prototypes (see~\cite{poldma2009learning}).

\subsection{Color Semantics in HCI}

Although some studies in HCI investigate the role of colors in aesthetics and usability~\cite{lavie2004assessing,hassenzahl2004interplay,hoffmann2004critical,de2006interaction,schmidt2009webpage,navarre2009icos,moshagen2010facets,lee2010effects,reinecke2011improving,sauer2011influence,reinecke2013predicting,thuring2007usability,lindgaard2007aesthetics,hassenzahl2010inference,lindgaard2011exploration,lee2011impact,yang2012deep} and performance of the user~\cite{moshagen2009blessing,sonderegger2010influence,linSemantically2013}, the notion of color semantics has been barely addressed.
As an example, consider a scenario in which
users are surfing the Web for a ``massage therapy'' website. Users perhaps expect to see
a website with color combinations that impress them with ``calm'', ``soothing'', and ``elegant''
moods. If the color design of such a website fails to convey these moods, it will then negatively affect the trust, credibility, and marketing of its business (See~\cite{lindgaard2011exploration,pengnate2013multimethod}).

We believe that color semantics is a legitimate challenge to be addressed in various venues of HCI: in automatic design~\cite{lok2001survey,gajos2004supple,hurst2009review,jahanian2012automatic,hunter2011web,Kuhna2012semi}, design by example~\cite{hartmann2008design,herring2009getting,lee2010designing,kumar2011bricolage}, design grammar~\cite{talton2012learning}, design of user interaction~\cite{carroll1997human,fallman2003design,lowgren2007pliability,zimmerman2007research,forlizzi2008crafting,stolterman2008nature,hashim2009design,chen2011interface}, quantifying aesthetics of design~\cite{lavie2004assessing,datta2006studying,reinecke2013predicting}, user experience design~\cite{hassenzahl2006user,mahlke2008visual,hartmann2008framing,lowgren2009toward,law2011measurability,bargas2011old,van2012user}, and color design~\cite{luo2006applying,tokumaru2002color,tsai2007automatic,wang2008color,ou2009colour,hsiao2013consultation,jahanian2013automatic}. 
As Fallman argues, HCI needs to be understood and acknowledged as a design-oriented process in terms of philosophy and theory and methodological foundations~\cite{fallman2003design}. Dearden and Finlay suggest that we need to identify patterns that are both timeless and culturally sensitive~\cite{dearden2006pattern}. Hoffmann and Krauss~\cite{hoffmann2004critical} conclude that many researchers do not recognize the importance of visual aesthetics in the communication intent.
We need to work on semantic color vocabularies and define and propose them as domain-specific knowledge to HCI.
This motivates us to revisit the concept of color semantics, utilize it in design mining and data-driven approaches to learn from the work of professional designers, and investigate theoretical and practical aspects of color semantics in user interaction design.

\section{Related Work}
\label{sec:RelatedWork}
Our work lies at the intersection of four research areas: color
semantics and meaning, probabilistic topic models, user study via
crowdsourcing, and click modeling.

\subsection{Color Semantics and Meaning}

Semantics in computer science and engineering has been observed as a
gap~\cite{liu2007survey,eakins1999content,sethi2001mining,mojsilovic2001capturing,zhou2000cbir,smeulders2000content,chen2003unsupervised}
that merits more investigation.  Nevertheless, in terms of color
emotions and moods, there is a recent body of work.  Csurka et
al.~\cite{csurka2010learning} collected a dataset of color combinations
and their associated labels, and applied a Gaussian mixture model to the
data based on some low level color features of the color combinations.
Unlike our work, their dataset is collected from a limited number of
good designs suggested in~\cite{eisemann2000pantone} and mainly
from an online dataset of colors~\cite{colourlovers} produced mostly by
amateurs.  The noisiness of the labels in this online dataset was
acknowledged by both these authors and~\cite{o2011color}.  Another
difference is that their model does not categorize the color
combinations as joint combinations of labels and colors.

Applications of color semantics have recently come to the attention of researchers.  In image retrieval, Solli and
Lenz~\cite{solli2010color} defined a mathematical framework for
Kobayashi's Color Image Scale.  This framework is utilized
by~\cite{jahanian2013recommendation,jahanian2013automatic} in designing
alternative and customized magazine covers for non-designers based on
color moods.  For color mood transfer of images, Murray et
al.~\cite{murray2012toward} applied the 15 moods suggested by Csurka et
al.~\cite{csurka2010learning}.
In contrast to our work, where we use human judgment to evaluate
the goodness of our model, they use human input to build their models.
In design mining the Web, Kumar et
al.~\cite{kumar2013webzeitgeist} reported using expressive colors (color
moods) to support data-driven design tools.

In researching linguistic categorization of colors, color naming and its
cross-cultural aspects are topics that have been under
investigation for decades~\cite{palmer1999vision}. There is a body of work in computer science and engineering to model color naming (e.g.~\cite{mojsilovic2005computational}.)
In recent work, Heer
and Stone~\cite{heer2012color} reviewed statistical color naming models,
with the goal of fitting a model to single colors and their associated
names.  These associations are either from human judgments or retrieved
from Internet search engines.  For the latter, topic modeling was
explored by Weijer et al.~\cite{van2009learning} who used Probabilistic
Latent Semantic Analysis (PLSA), and Schauerte and
Stiefelhagen~\cite{schauerte2012learning} who used Latent Dirichlet
Allocation (LDA).  In contrast to color naming, our color semantics work
takes into consideration design examples, color combinations, and
different levels of abstraction.  Also, unlike prior work, we adapt
topic models to jointly link colors and words, and then verify the
associations through user studies.

\subsection{LDA Topic Modeling}

The goal in LDA topic modeling is to infer underlying themes or topics
of textual document corpora, where each topic is a multinomial
distribution over words, and each document is a mixture of
topics~\cite{blei2003latent,Blei:2012:PTM}.
Because for color semantics we need to infer compound (color-word) topics, we adapt LDA-dual~\cite{shu2009latent} by viewing each magazine cover as a combined bag of words
and colors.  The closest work to our research is perhaps
\cite{feng2010topic}, where LDA is used for image annotation.  The authors
consider a document as a mixture of low level image features (extracted
by SIFT techniques) and words.  However, in our adaptation of LDA-dual model, we
incorporate two independent multinomial distributions, one each for observed colors and words.

\subsection{User Studies via Crowdsourcing}

Viability of crowdsourcing graphical perception, including color
visualization, has been confirmed by recent work of Heer and
Bostock~\cite{heer2010crowdsourcing}.  In several tests, they replicated
the results of prior laboratory experiments using crowdsourcing and
reported consistency between the results.  The scalability yet
inexpensiveness of online experiments involving color overrides some of their
limitations, such as the inability to control for variations in visual acuity.  For instance, Lin et
al.~\cite{linSemantically2013} employ crowdsourcing to confirm
improvement in subjects' performance when color semantics is utilized in
information visualization.  Reineke et
al.~\cite{reinecke2013predicting,reinecke2014} utilize cross-cultural
crowdsourcing to study some perceived aesthetics aspects of visual
complexity and colorfulness in a dataset of website designs.  Inspired
by their online setup, we also captured demographic information from the subjects during our
crowdsourcing experiments. However, we emphasize that our goals from
the crowdsourcing experiments are completely different from work of Reinecke et al.  While they wish
to infer bias in perception based on demography and cultures, we use this information to validate the output of a statistical model.

\subsection{Click Modeling}

The goal in click modeling is to model the user's interactions with
sponsored search results or ads.  In a recent work, Govindaraj et
al.~\cite{govindaraj2014modeling} reviewed click models, and suggested a
new model by taking into account relations between the user's clicks on
a list of URLs (provided by a search engine in response to the user's
query).  Modeling the user's back and forth clicks on a list of URLs,
they then infer the probability of clicking on different vertical
positions of URLs, regardless of their contents.  Inspired by this work,
we treat the participants responses as a click modeling problem.
Arguably our model is simpler than that proposed by Govindaraj et al. However, unlike web-search where the user can choose not to click on
any of the results, in our case the user needs to make a
choice. Furthermore, since our users participate willingly in our study, the level of user engagement is high. Therefore, we believe
that our simple model suffices.

\section{Data Collection}
\label{sec:DataCollection}

Our dataset of magazine covers includes 2,654 covers from 71 magazine
titles and 12 genres.  We collected approximately 1,500 of these covers by scanning
them from magazines held by a number of libraries and newsstands in our
university.  The rest of the cover images were downloaded from the
Internet.  Although we developed a web crawler tool to collect magazine
covers, because many magazine publishers do not provide archives with
high quality images, in half of the cases we had to collect online
images by hand\footnote{This data will be made available for academic research
upon request.}. We attempted to collect roughly 12 different genres of magazines to capture different contexts of design.  These genres include \emph{Art}, \emph{Business}, \emph{Education}, \emph{Entertainment}, \emph{Family}, \emph{Fashion}, \emph{Health}, \emph{Nature}, \emph{Politics}, \emph{Science}, \emph{Sports}, and \emph{Technology}. To this
end, we loosely followed the Dewey Classification method~\cite{Dewey}, the WorldCat indexing system~\cite{worldcat},
suggestions from our librarians, as well as the description of the
magazine by the publishers.  Table \Rmnum{1} in the appendix contains a summary of our dataset.  Note that the genres are tentative.
In fact, this supports the use of LDA: magazines rarely include only one
topic and usually are a combination of different topics (or genres in
this context).

\subsection{Preprocessing}
\label{sec:Preprocessing}

The preprocessing of cover images was performed using the Matlab \emph{Image Processing} toolbox\footnote{The MathWorks, Inc., Natick, MA.}.
We use 512 basic colors obtained by quantizing the sRGB color space with 8 bins in
each channel. Given this color basis, each magazine cover (image) is then a
histogram of these colors.  To feed the images to LDA-dual, we scale them
to 300$\times$200 pixels using bicubic interpolation.
The down-sizing was done to reduce the computation without affecting the distribution of the colors in the images.

To extract color palettes from the images, we used the color theme extraction code by~\cite{2013-color-themes}. In their implementation, the algorithm needs to take the saliency map of the given image (using the code by~\cite{judd2009learning}), as well as the segmentation of the given image (using the code by~\cite{felzenszwalb2004efficient}). In our work, we however used the saliency map code by~\cite{harel2007graph}, since it was easily accessible. Note
that these color palettes are not the input to the model, but are later used to visualize the inferred
color combinations.

\subsection{Word Vocabulary}

To capture the words to be associated with color distributions of the magazine
covers, the words on the covers were transcribed by hand. To create a
word vocabulary, we first prune the transcribed words and then create a
histogram of words. Because a more meaningful vocabulary results in more
meaningful topics, we filter out special characters, numbers, common stop
words\footnote{Provided by MySQL database, available at https://dev.mysql.com/doc/refman/5.1/en/fulltext-stopwords.html.} (e.g.\ articles and lexical words), and an additional handcrafted list of stop words (see Table \Rmnum{2} in the appendix).
Compound words formed with a hyphen or dash are
decomposed.  A version of the Porter Stemming
algorithm~\cite{porter1980algorithm} is used to equate different forms
of a word, for instance ``elegant'' and ``elegance.'' Finally, a mapping
from month to season is applied.  In order to include the context and
classes of the magazines with the associated words, the periodical category to which
each magazine title belongs was added to the set of words. We
collected these periodical categories from the WorldCat indexing system,
which is the largest international network of library content and
services~\cite{worldcat}.

\section{Statistical Model}
\label{sec:StatisticalModel}

\begin{figure}[h!tb]\tiny
  \centering
  \subfloat[][]{ \includegraphics[scale=.5]{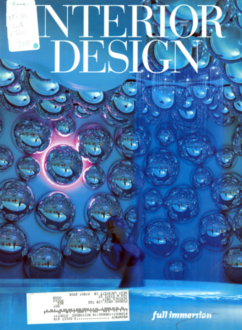}} \ \
  \subfloat[][]{ \includegraphics[scale=.5]{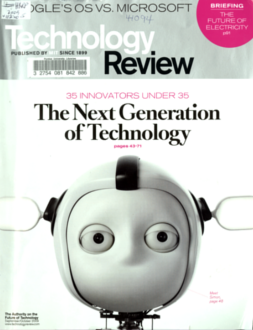}} \ \
  \subfloat[][]{ \includegraphics[scale=.5]{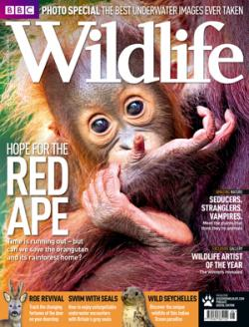}} \ \
  \caption{Pink is used in all of these designs, despite the fact that each of these designs belongs to a different context and genre of magazines.}\label{PinkInDesign}
\end{figure}

When ideating about visual design, the designer takes into account the topic or
the context within which he or she is asked to convey his or her message.  For
instance, when the context is \emph{politics}, the designer may tend to use darker, ``heavier'' and more ``formal'' colors.
However this is not the only factor, the words in the design also
influence the designer's choice of colors. Figure~\ref{PinkInDesign}
illustrates that \emph{pink}\,---\,which is usually used for topics related to feminity\,---\,has been used in a variety of magazines from
different genres.  This observation suggests that each design's
theme might be a combination of words and color distributions; and each
design may include a proportion of various themes.
Our goal is to model these combinations of words and colors, and infer proportions of these combinations in magazine cover designs.
A similar intuition
has been argued in statistical topic modeling, specifically
LDA~\cite{blei2003latent}, for modeling word distributions in documents as proportions of different word topics.

LDA (Latent Dirichlet Allocation) is an intuitive approach to infer
topics from text data.  As Blei et
al.~\cite{blei2003latent,Blei:2012:PTM} describe, instead of
categorizing and exploring documents using tools such as keywords, we
may first categorize documents based on topics.  This allows us to
explore topics of interest and find related documents.  For example, a
document about sociology may include different topics, such as biology,
evolution, history, and statistics with different proportions.  Each of
these individual topics can be viewed as a multinomial distribution over
a fixed vocabulary of words. Accordingly, each document, which can be
viewed as a bag of words, is a combination of these topics with some
proportions.  Typically, a value for the number of topics is chosen by
hand. The latent topics, as well as the topic proportions of each
document, are inferred by LDA using the observed data, which are the words
in the documents.

Just as word topics are distributions over words, one may think of color topics as distributions over colors. One may argue that by only using LDA, we might be able to model the distribution of the colors on the covers, and infer the color topics. We could then utilize crowdsourcing, and ask participants to label the inferred color topics to include combined color-word topics, and hence the color semantics associations.
The problem with this approach is that we cannot capture the associations created in each design by the designer. Moreover, we cannot perform this modeling in an automatic fashion, since we need human judgements to label the color topics, and this labeling will be performed by naive subjects.
Hence, we will lose the valuable insight that the designer put into selecting color and word topic combinations.

A better approach is to model the links or the associations between the color topics and word topics to infer combined color-word topics.
This means that the LDA model framework needs to be extended. Such an extension was recently proposed by Shu et al.~\cite{shu2009latent}. The key to this extension is the proportions vector $\boldsymbol{\theta}$.
In our case, for each cover design, the proportion is a combination of color assignments, as well as word assignments.
Similar to LDA, these assignments are modeled with multinomial (and conjugated by Dirichlet) distributions.

\subsection{Review of LDA-dual Model for Color Semantics}

In this section, we review the LDA-dual model proposed by Shu et. al~\cite{shu2009latent}, and explain how to adapt this model for color semantics.
Our implementation\footnote{Our implementation is available upon request.} is an extension of the Matlab \emph{Topic Modeling} toolbox~\cite{MatlabTopicModelingToolbox} developed by~\cite{griffiths2004finding} for LDA.

Assume that there are $K$ color-word topics denoted by $k_1, k_2,..., k_K$
and $D$ magazine covers denoted by $d_1, d_2, ..., d_D$.  Let $W$
denote the number of words in the vocabulary and $C$ denote the number
of color swatches, where each swatch is a patch of color defined by
using its sRGB values\footnote{Recall that we discretize and use eight
  values for each of the three sRGB color channels. Therefore, $C =
  512$.}.  Moreover, let $M_{d}$ denote the number of words and $N_{d}$
denote the number of color swatches in magazine cover $d_{d}$. Let
$w_{d,m}$ denote the $m$-th word in the $d$-th document and $c_{d,n}$
denote the $n$-th color swatch in the $d$-th document.
Each magazine cover includes some proportion of each word topic, as well as each color topic.
Let $y_{d,m}$ denote the word topic assignment to the word $w_{d,m}$ and $z_{d,n}$ denote the color topic assignment to the color swatch $c_{d,n}$. Note that these assignments are latent. Also let $\psi_{y_{d,m}}$ and $\phi_{z_{d,n}}$ denote the multinomial distributions of the word topics and the color topics, respectively.

Each magazine cover includes some proportion of the color-word topics.  These
proportions are latent, and one may use the $K$ dimensional probability
vector $\theta_{d}$ to denote the corresponding multinomial distribution
for a document $d_{d}$.

Let $\beta$, $\gamma$, and $\alpha$ be the hyper-parameters of the three Dirichlet distributions for the color topics, word topics, and the proportions $\theta_{d}$, respectively. Let $\mathtt{Dirichlet}(\cdot)$ denote the Dirichlet distribution, and $\mathtt{Discrete}(\mathtt{Dirichlet}(\cdot))$ denote the discrete distribution that is drawn from a Dirichlet distribution.

Given the above notation, the generative model for LDA-dual can be written
as follows:
\begin{enumerate}
\item Draw $K$ word topics $\psi_{k} \sim \mathtt{Dirichlet}(\gamma)$.
\item Draw $K$ color topics $\phi_{k} \sim \mathtt{Dirichlet}(\beta)$.
\item For each document $d_{d} \in \{d_1, d_2, \ldots, d_D\}$:
  \begin{itemize}
  \item Draw $\theta_{d} \sim \mathtt{Dirichlet}(\alpha)$.
  \item For each word $w_{d,m}$ with $m = 1, \ldots, M_{d}$
    \begin{itemize}
    \item Draw $y_{d,m} \sim \mathtt{Discrete}(\theta_{d})$
    \item Draw $w_{d,m} \sim \mathtt{Discrete}(\psi_{y_{d,m}})$
    \end{itemize}
  \item For each color $c_{d,n}$ with $n = 1, \ldots, N_{d}$
    \begin{itemize}
    \item Draw $z_{d,n} \sim \mathtt{Discrete}(\theta_{d})$
    \item Draw $c_{d,n} \sim \mathtt{Discrete}(\phi_{z_{d,n}})$
    \end{itemize}
  \end{itemize}
\end{enumerate}

\begin{figure}[h!tb]
  \centering
  \includegraphics[width=0.6\textwidth]{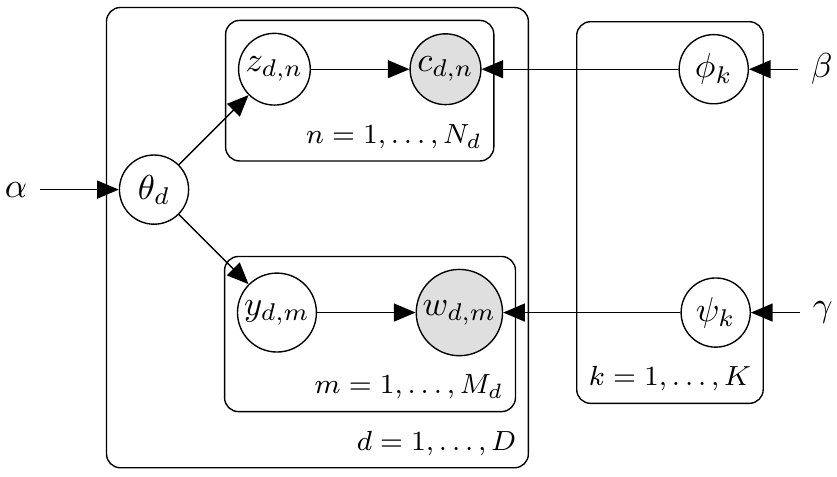}\caption{Hierarchical Bayesian plate model for the LDA-dual model, which
  combines color and word topics. Here, $D$ is the number of magazine covers;
  $K$ is the number of color-word topics; and $N_d$ and $M_d$ are the number
  of color swatches and words, respectively, in the $d$-th magazine cover.}
\label{LDAGraph}
\end{figure}

A graphical model for this generative process is illustrated in
Fig.~\ref{LDAGraph}, where the shaded nodes denote observed random
variables and the unshaded nodes are latent random variables.

If we let $\boldsymbol{\phi} = \{ \phi_{1}, \ldots, \phi_{K}\}$,
$\boldsymbol{\psi} = \{ \psi_{1}, \ldots, \psi_{K}\}$,
$\boldsymbol{\theta} = \{\theta_1, \ldots, \theta_{D}\}$, $z_{d} =
\{z_{d,1}, \ldots, z_{d, N_{d}}\}$, $y_{d} = \{y_{d,1}, \ldots, y_{d,
  M_{d}}\}$, $\boldsymbol{z} = \{ z_{1}, \ldots, z_{D} \}$,
$\boldsymbol{y} = \{ y_{1}, \ldots, y_{D} \}$, $\boldsymbol{w} = \{ w_1,
\ldots, w_{d}\}$, and $\boldsymbol{c} = \{ c_1, \ldots, c_{d}\}$, then
the joint distribution corresponding to the LDA-dual model above can be
written as
\begin{align}
  \nonumber p \left( \boldsymbol{\phi},\boldsymbol{\psi},
    \boldsymbol{\theta},\boldsymbol{z}, \boldsymbol{c},
    \boldsymbol{y},\boldsymbol{w} \right) =
  \nonumber & \prod_{i=1}^{K} p \left( \phi_{i} | \beta \right) \cdot p \left(
    \psi_{i} | \gamma \right) \cdot \prod_{d=1}^{D}{p \left( \theta_{d} |
      \alpha \right)}    \cdot \left( \prod_{n=1}^{N} { p \left( z_{d,n} | \theta_{d}
      \right) p \left( c_{d,n} | \boldsymbol{\phi}, z_{d,n}
      \right)} \right) \\
  & \cdot \left( \prod_{m=1}^{M} { p \left( y_{d,m} | \theta_{d} \right) p
      \left( w_{d,m} | \boldsymbol{\psi},y_{d,m} \right)} \right).
    \label{eq:lda}
\end{align}

\begin{figure}[h!tb]
  \centering
  \includegraphics[width=1\textwidth]{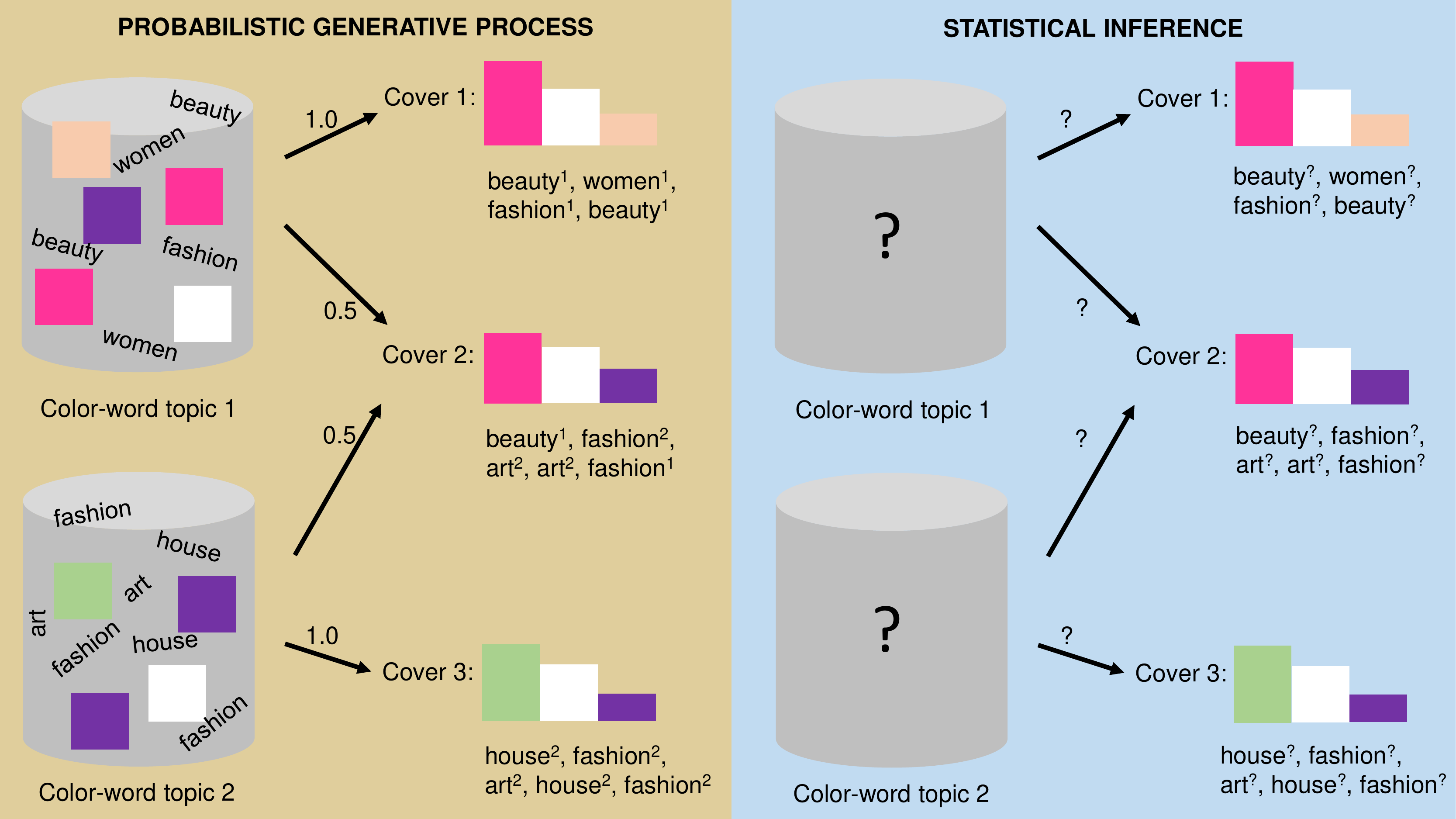}
  \caption{LDA is both a generative and inference model. This image is inspired by \protect\cite{steyvers2007probabilistic}.}
  \label{LDAGenerative}
\end{figure}

Figure~\ref{LDAGenerative} provides a graphical illustration of the
generative mechanism and the inference procedure described below.
This figure is a symbolic representation of the model. In each sub-figure, a cylinder represents a color-word topic. Each arrow represents the probability of each cover being drawn from a given color-word topic. Each cover includes a histogram of colors and a list of words (each word is superscripted by its corresponding color-word topic). In the generative process, we know the distribution of the color-word topics, and can produce the distribution of the colors and words on the magazine covers. For instance, ``Cover 1'' is completely (with probability $1.0$) generated by color-word topic 1. ``Cover 2'' is generated by equal distributions of both ``color-word topic 1'' and ``color-word topic 2''. In the statistical inference mechanism, we only know the distribution of the colors and the words for each cover. We do not know (represented by question marks) the color-word topics, their proportions, and the assignments of the colors and words of each cover to these color-word topics.

\subsection{Inference}
\label{sec:Inference}

Since $\boldsymbol{c}$ and $\boldsymbol{w}$ are observed, inference
entails computing

\begin{align}
  \label{eq:lda-inference}
  p\left( \boldsymbol{\phi}, \boldsymbol{\psi},
    \boldsymbol{\theta},\boldsymbol{z},
    \boldsymbol{y}|\boldsymbol{c},\boldsymbol{w} \right) = \frac{ p
    \left( \boldsymbol{\phi},\boldsymbol{\psi},\boldsymbol{\theta},
      \boldsymbol{z},\boldsymbol{y},\boldsymbol{c},\boldsymbol{w}
    \right)} {p \left( \boldsymbol{c},\boldsymbol{w} \right)}.
\end{align}

Theoretically, the above distribution can be obtained by computing the
joint probability distribution of the latent and the observed variables,
and then computing the marginal probability of the observations. In
practice, however, topic modeling algorithms approximate the result to
bypass the computational complexity of the solution. There are often two
approaches for this approximation~\cite{Blei:2012:PTM}: variational inference~\cite{jordan1999introduction,teh2006collapsed} and Markov chain Monte Carlo (MCMC) sampling~\cite{andrieu2003introduction,griffiths2002gibbs}.
For the sampling approach, the collapsed Gibbs version of sampling is discussed in~\cite{porteous2008fast,griffiths2004finding,shu2009latent}.
We extend MCMC collapsed Gibbs sampling version implemented in the Matlab Topic Modeling
toolbox~\cite{MatlabTopicModelingToolbox} developed by~\cite{griffiths2004finding} for LDA (see~\cite{jahanian2014quantifyingaesthetics} for our derivations of LDA-dual).

\begin{figure*}[h!tb]
  \centering
  \subfloat[][]{ \includegraphics[scale=.6]{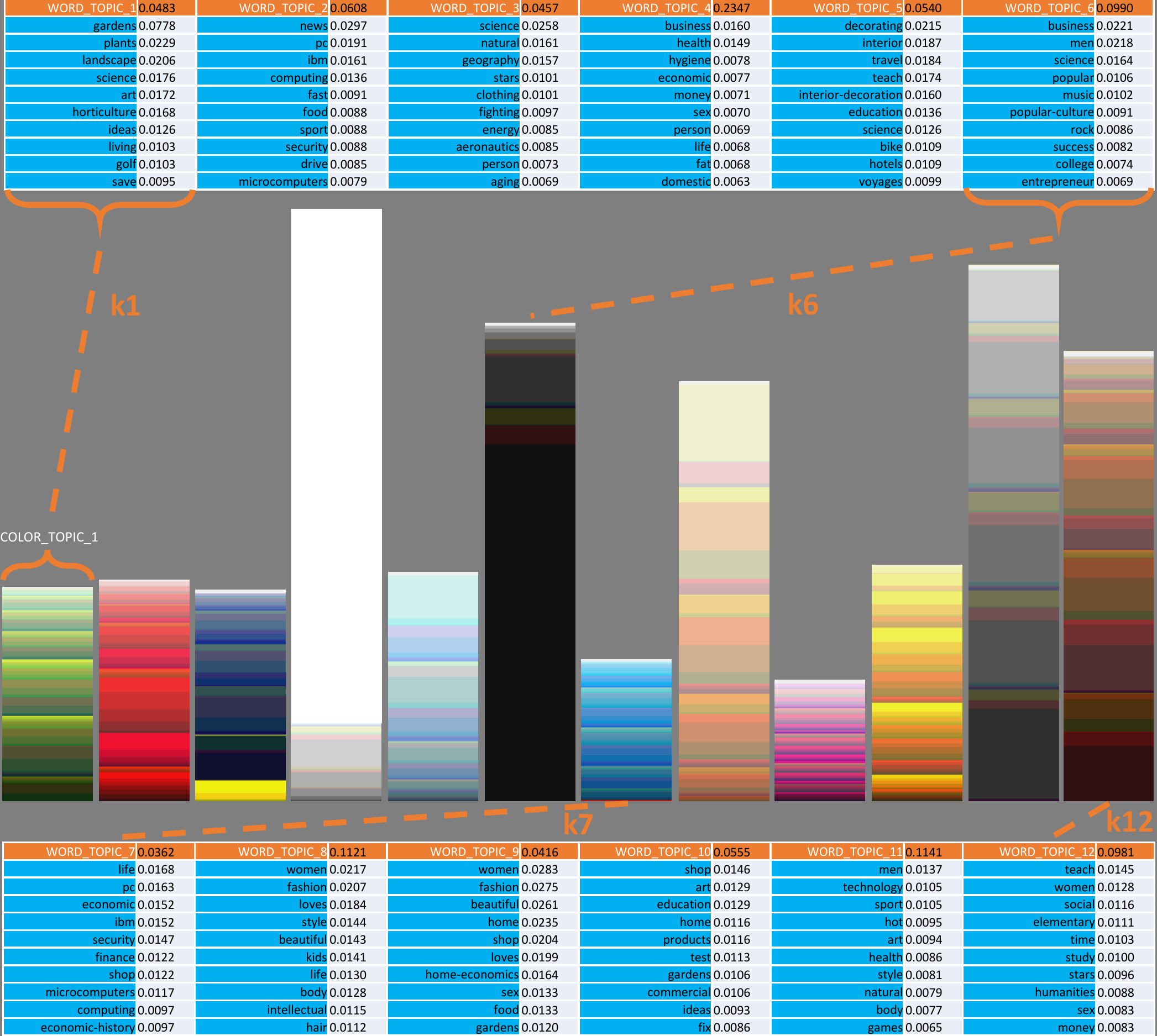}}\\
  \subfloat[][]{ \includegraphics[scale=.33]{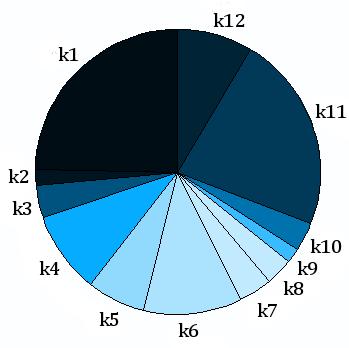}} \ \ \ \ \ \ \ \ \ \
  \subfloat[][]{ \includegraphics[scale=.33]{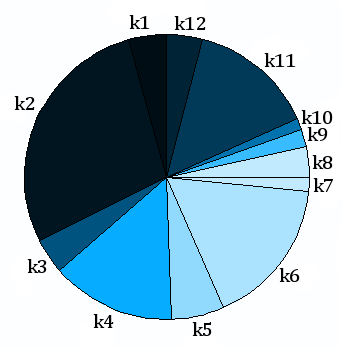}}  \ \ \ \ \ \ \ \ \ \
  \subfloat[][]{ \includegraphics[scale=.33]{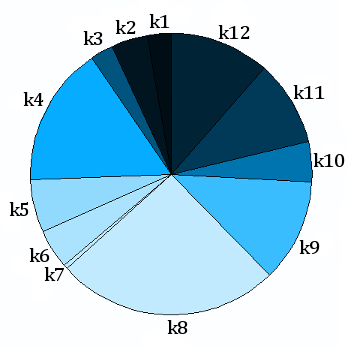}}
  \caption{Color-word topics inferred by the LDA-dual model.  (a)
    Illustration of the 12 color topics in the middle, and their
    corresponding 12 word topics; 6 on top for the first 6 color
    histograms from left, and the other 6 on the bottom.  Note that for
    visualization, only the principal elements in the histograms are
    shown.  Also note that the numerical weight of each word topic is
    shown next to heading of each word topic histogram.  Similarly, the
    numerical weights of the words in each word topic histogram are
    shown next to their corresponding words (see Sec.~\ref{sec:Inference}). As examples, the proportions of the 12
    color-word topics in (b) \emph{Horticulture}, (c) \emph{Time}, and
    (d) \emph{Vogue} magazines (including all the issues in the dataset from year 2000 to 2013) are shown. Note that $k$'s are the same as in (a). See Fig.~\ref{topicsVsTitles} for all of the magazines.}
  \label{color-word-histograms}
\end{figure*}

Although some variants of LDA can automatically find an optimal value for the number of topics $K$,
based on our domain knowledge from the data collection process, we
simply set $K=12$. Because each color-word topic includes proportions of
the color basis and the vocabulary words, we visualize a topic as a pair of
colors and words histograms.  Figure~\ref{color-word-histograms} (a)
illustrates the 12 topic histogram pairs for $K=12$, $\alpha=0.8$, and
$\beta=\gamma=0.1$.
We tuned the values for $\alpha$, $\beta$, and $\gamma$ by trial and error to produce visually pleasing color
histograms and semantically meaningful word histograms.
The visualized histograms just illustrate the principal components.  Of course, other levels of granularity can be visualized if needed (see~\cite{jahanian2014quantifyingaesthetics} for more examples).
Note that in this figure, similar to ``WORD\_TOPIC\_1'', ``COLOR\_TOPIC\_1'' is a distribution over some colors. WORD\_TOPIC\_1 itself has some proportion in the entire dataset ($0.0483$). The summation of all the $12$ word topics proportions is $1$. Similarly, all the $12$ color topic proportions add up to $1$. Here, just for visualization, we show the length of each color topic based on its ratio to the color topic ---COLOR\_TOPIC\_4, which has the largest proportion.

\begin{figure}[H]
  \centering
  \includegraphics[width=0.7\textwidth]{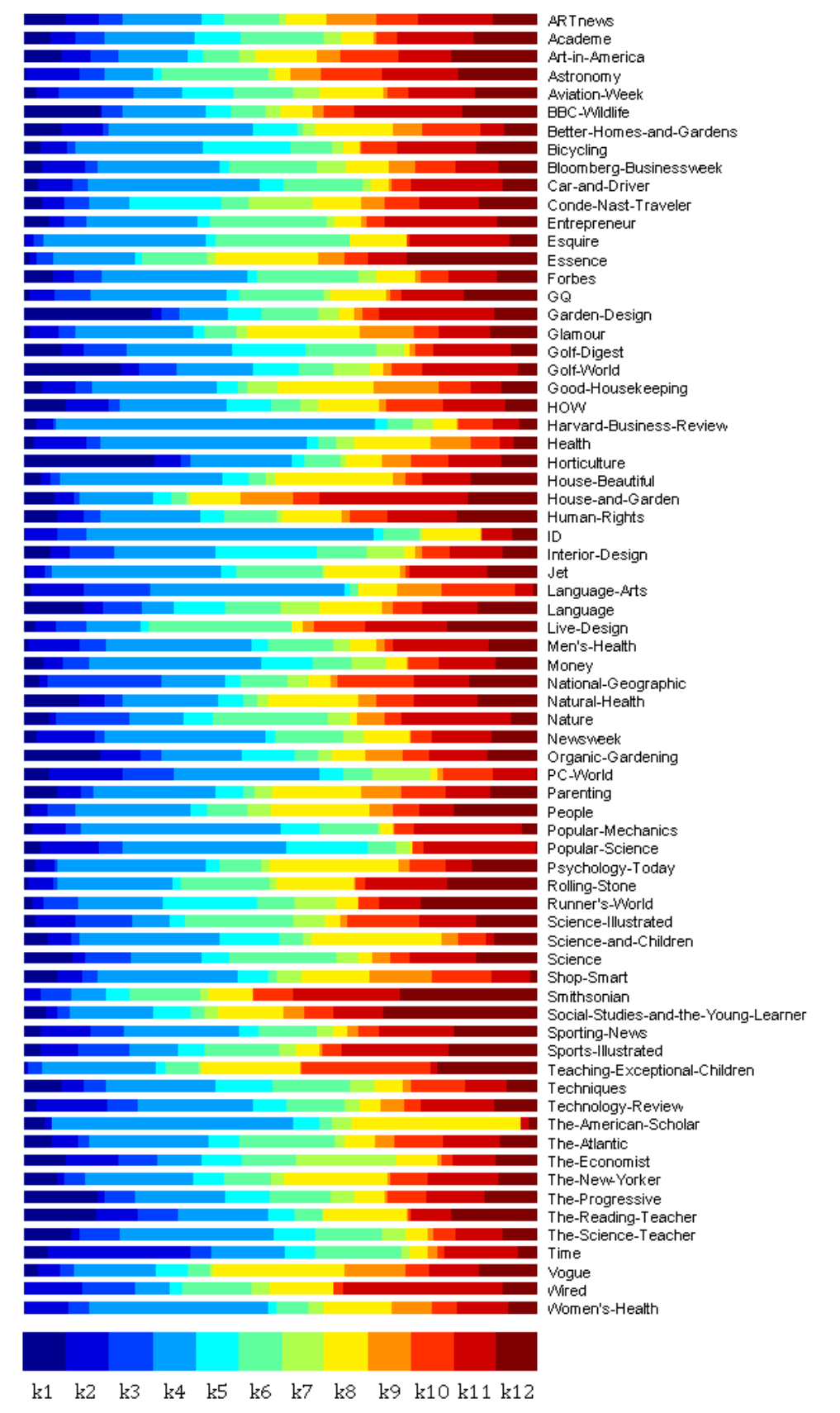}
  \caption{Topics vs. titles. The proportion of each of the 12
    color-word topics, $k_1$ to $k_{12}$ (see Fig.~\ref{color-word-histograms}) for each
    magazine title including all the issues in the dataset from year 2000 to 2013 is illustrated. Note that the colors here are just
    legends for the purpose of visualization, and are not related to the
    color-word topics. }
  \label{topicsVsTitles}
\end{figure}

Figures~\ref{color-word-histograms} (b), (c), and (d) illustrate the
proportions of each of the inferred color-word topics for three magazine
title designs in the dataset.  For instance, note that \emph{Vogue} as a
fashion magazine has $k_8$ and $k_9$ as two of the dominant color-word
topics. As can be seen, $k_8$ and $k_9$ contain words such as
``women'', ``fashion'', ``love'', and ``beauty'', while the
corresponding color histograms contain pastel and pink colors, which are
often associated with fashion magazines. On the other hand,
\emph{Horticulture}, which is a nature magazine, has the highest
proportion of $k_1$, which pre-dominantly contains shades of green. The
words in $k_1$ include gardening-related words such as ``gardens'',
``landscapes'', and ``plants''. See Fig.~\ref{topicsVsTitles} for all 71 magazine titles.

\section{Interpreting The Model Output}
\label{sec:InterpretingModelOutput}

Visualizing the results of LDA is a topic of research~\cite{chaney2012visualizing,chuang2012termite}.  Chaney and
Blei~\cite{chaney2012visualizing}, for example, suggest a visualization
mechanism for exploring and navigating through inferred topics from LDA
and their corresponding documents.  Although their work does not completely address the usability evaluation of
this mechanism, it inspired
our visualization mechanism for our user study.  In order to evaluate
the color semantics hypothesis, we need to display both the color
histogram and the word histogram to the participants in our user study
in a comprehensive, yet unbiased fashion.  We address this via a two-step
process. The word histograms are converted to word clouds, while the
color histograms are converted to color palettes using the mechanism
described below. Figure~\ref{VisProcesses} illustrates the
visualization process.

\subsection{From Color Histograms to Color Palettes}

We use 5-color palettes as proxies to represent each of the color
histograms in Fig.~\ref{color-word-histograms} (a). These palettes
are drawn from a pool of 5-color palettes, one for each magazine cover in the dataset (refer to Sec.~\ref{sec:Preprocessing} for the mechanism of creation of this pool).

The intuition behind using 5-color palettes is that when designers are asked
to design a media piece, they often choose a 3-color, 5-color,
or occasionally a 7-color palette, and consistently use it; so
that their designs are clean and sophisticated, as opposed to
busy and cluttered.

\begin{figure}[h!tb]
  \centering
  \includegraphics[width=1\textwidth]{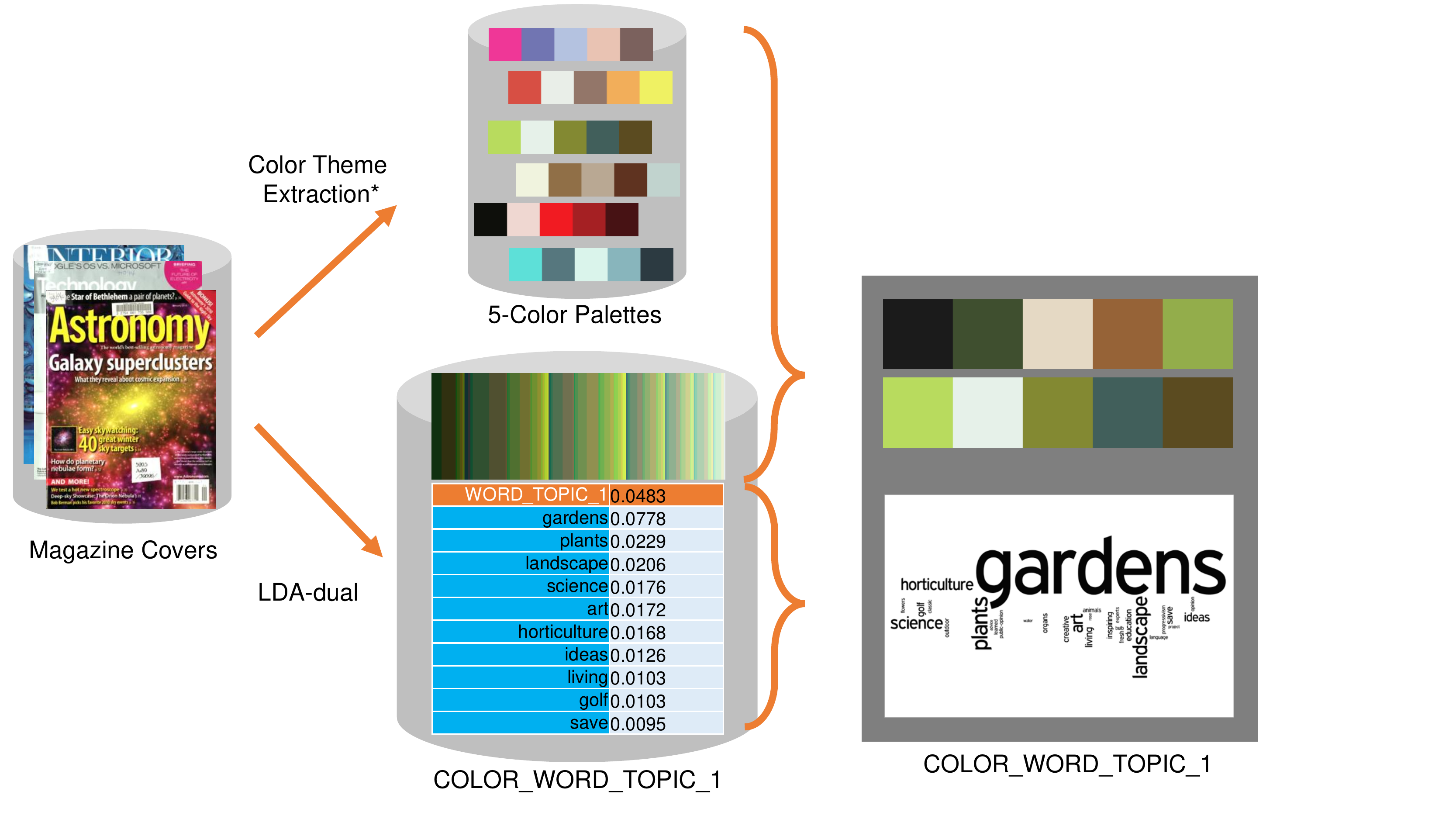}
  \caption{Visualization process for the inferred color-word topics. To
    visualize the color-word topic histograms inferred by the model (see
    Fig.~\ref{color-word-histograms}), we use 5-color palettes and
    word clouds as proxies to color histograms and word histograms,
    respectively. *See Sec.~\ref{sec:Preprocessing} for the color theme extraction details.}
  \label{VisProcesses}
\end{figure}

In order to find the 5-color palettes that are closest to the color topic
histograms, we define a similarity metric as follows: Let $S^{512}$
denote a color topic histogram with the 512 color basis defined earlier, and $S^{5}$
denote a 5-color palette.  An intuitive similarity metric is the Euclidean
distance between color swatches of $S^{512}$ and $S^{5}$.  Among the
possible color spaces, we choose the CIE L*a*b* color space with a D65 reference white
point. It is considered to be a perceptually uniform space, where $\Delta E$ around $2.3$ (the distance between two colors) corresponds to one JND (Just Noticeable
Difference)~\cite{sharma2002digital:ch1}.

Defining the color similarity distance problem as a bipartite graph
matching between $S^{512}$ and $S^{5}$ with $512$ and $5$ nodes,
respectively, we find the minimum distance cost of this graph using the
Hungarian method~\cite{kuhn1955hungarian}.  Equation~\ref{eq3} defines
the weighted Euclidean distances $d_{WED}$ between the nodes of these two graphs.  Here, the
weight $w_i$ corresponds to the weight of the $i$-th color in the color
topic histogram $S^{512}$; and $\left \| S^{512}_{i} - S^{5}_{j} \right \|_2$ denotes the distance in CIE L*a*b* between the $i$-th color from $S^{512}$ and the $j$-th color from $S^5$.
This metric can be thought as a version of
The Earth Mover's distance suggested by Rubner et
al.~\cite{rubner2000earth} for image retrieval, with the weight vector representing color importance.
\begin{align}
  \label{eq3}
  d_{WED} = \sum_{i=1}^{512}{\frac{1}{w_i}\sum_{j=1}^{5}{\left \|
        S^{512}_{i} - S^{5}_{j} \right \|_2}}.
\end{align}

Computing $d_{WED}$ for a given color topic histogram and all 5-color
palettes, we can choose the closest of them as proxies to the
histogram (see Fig.~\ref{VisProcesses}).  In the user study, we
present two series of questions for the first and the second closest
color palettes, because just one color palette may not provide an adequate visualization of the entire topic histogram. See Fig.~\ref{color-word-topics} for the entire color-word topics.

\subsection{From Weighted Bag of Words to Word Clouds}

Figures~\ref{VisProcesses} illustrates how we visualize each
word topic histogram with a word cloud.  The word cloud (or tag cloud) is a
visualization technique used to show the relative weights of words through
different font sizes.  The weights resemble frequency of occurrence or
importance of the words in a word dataset.  A suite of word cloud
algorithms and their usabilities are discussed in~\cite{seifert2008beauty}.  Because of the popularity of word clouds
in visualizing categories, and the fact that words are randomly
scattered over a layout, we used this technique in our user
study.  Using wordle\footnote{http://www.wordle.net}, we generated black
and white word clouds to avoid introducing any color bias.
Note that whereas we chose two color palettes for each color topic, we developed only one word cloud for each word topic.

\section{User Study}
\label{sec:UserStudy}

The main aim of this section is to validate the output of the
probabilistic topic model. In particular, we want to understand
if casual users (who are not necessarily designers) agree
with the association between color combinations and linguistic
concepts produced by our model.

\subsection{Participants}

Our color semantics online survey\footnote{http://goo.gl/P4W9XL} is
hosted at our university's survey system.
It was advertised through social networks and the university email
network.  So far, we have collected 859 responses from 487 (56.69\%)
females, 367 (42.72\%) males, and 5 others (0.58\%), in the age range of
18 to 80 years (with mean $=30.98$).  The participants are from 70 countries and natively
speak 66 different languages, with the majority from the US (59.84\%).  There are
348 (40.51\%) participants who have lived in more than one country.
There are 352 (40.97\%) participants with college degrees, 451 (52.50\%)
with graduate degrees, and 55 others (pre-high school, high school, and
professional degree).  The majority of the participants, 716 (83.35\%)
are non-designers.  In contrast, there are 130 (15.13\%) participants
with three or more years of experience in visual design (including
graphic design, interior design, and textiles.)

   \begin{figure}[h!tb]
     \centering
   \includegraphics[width=1\textwidth]{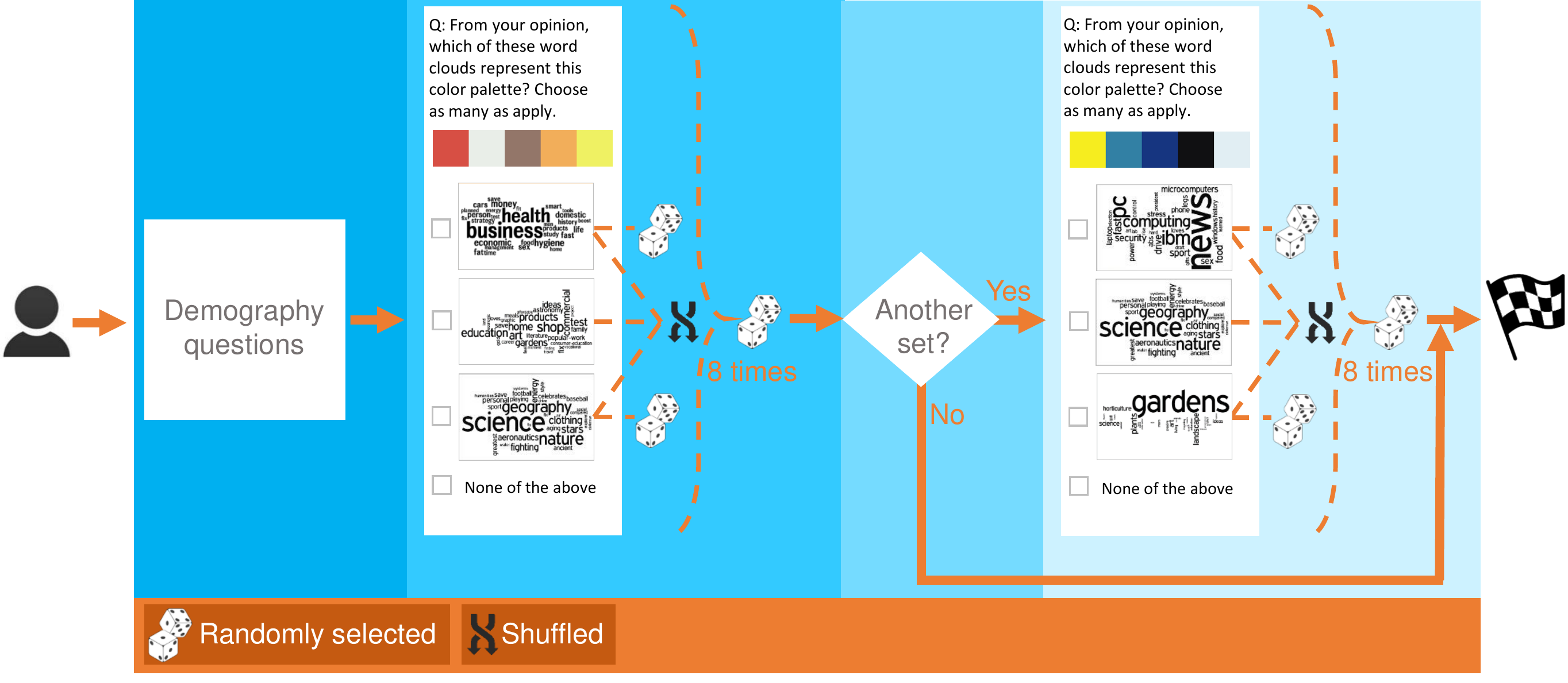}
 \caption{Flow of the user study.} \label{survey_flow}
   \end{figure}

\subsection{Stimuli and Procedure}

Figure~\ref{survey_flow} illustrates the flow of the survey.  In order
to simulate a matching experiment between pairs of color and word
topics, we designed a question as follows: one 5-color palette is shown
in the left side of the screen, and three shuffled and randomly chosen word
clouds, as well as a ``None of the above'' option are shown on the right side arranged in vertical order.
Each question is a multiple choice (represented by multiple choice check
boxes). For each question, we asked the participant to choose as many word clouds, as in his/her opinion apply to the 5-color palette shown. If the participant felt that none of the word clouds applied to the 5-color palette, he/she could choose ``None of the above''.
Among the three randomly drawn word clouds, one is the word cloud inferred from the model.

The survey is divided into two subsets of questions.
The reason for this is that otherwise some participants may lose interest in finishing the experiment.
This observation was made in the pilot experiment~\footnote{This pilot experiment is hosted at http://goo.gl/WFnjHL. The data collected for this experiment is available upon request.}.
More specifically, we created 24 questions for the first and second closest 5-color
palettes corresponding to the 12 inferred color topics.  However to
avoid exhausting the participants, we randomly draw 8 questions from the 12 questions of the closest color palettes and ask the participants to answer them.  Then we ask the participants if they
would like to continue by taking another set of 8 questions (this time drawn from the 12 questions of the second closest color palettes). Of all participants, 61.35\% of the users chose to continue, and answered all 16 questions.

Note that the reason for shuffling the word clouds in each question is because we
conjectured that there might be a position bias in the vertically arranged
options.  In Sec.~\ref{sec:InterpretingTheUserStudy}, we explain how we were inspired by
the click model in~\cite{govindaraj2014modeling}, and utilized this
framework to analyze the experimental results.

\section{Interpreting the User Study}
\label{sec:InterpretingTheUserStudy}
In this section we explain the statistical inference mechanism that we
used to understand the user responses.

\subsection{Statistical Model}

First, we define some notation. Let $c_{i}$ denote the event that the
$i$-th color palette was displayed. Also, let $w_{j}$ denote the
event that the user selected (clicked on) the $j$-th word cloud, and
$u_{ij}$ denote the probability that the $j$-th word cloud was selected by the user in response to the $i$-th color palette.
In order to compute $u_{ij}$ we note that
\begin{align}
  \label{eq:uij-def}
  u_{ij} & = \text{Pr}(w_{j} | c_{i}).
\end{align}
There are three possible positions $p \in \{1, 2, 3\}$ at which a word
cloud can be displayed. Let $d_{jp}$ denote the event that the $j$-th
word cloud was displayed at position $p$, and let $w_{jp}$ denote the
event that the user selected the $j$-th word cloud which was displayed
at the $p$-th position. Then
\begin{align}
  \label{eq:uij-def1}
  u_{ij} & = \sum_{p \in \{1, 2, 3\}} \text{Pr}(w_{jp} | d_{jp}, c_{i})
  \cdot \text{Pr} (d_{jp} | c_{i}).
\end{align}
If $d_{j}$ denotes the event that the $j$-th word cloud was selected
for display and $d_{j}^{p}$ the event that it was displayed at position $p$,
then
\begin{align}
  \label{eq:dp-def}
  \text{Pr} (d_{jp} | c_{i}) = \text{Pr} (d_{j} | c_{i}) \cdot \text{Pr}
  (d_{j}^{p} | c_{i}).
\end{align}
According to our experimental design, each word cloud has an equal
probability of appearing in any one of the three positions. Therefore
\begin{align}
  \label{eq:dp-def1}
  \text{Pr} (d_{j}^{p} | c_{i}) = \frac{1}{3}.
\end{align}
On the other hand, we always select the $i$-th word cloud (the true word
cloud according to our model) for the $i$-th color palette. The other two
slots are filled by selecting any two of the remaining 11 word clouds
uniformly at random. Therefore
\begin{align}
  \label{eq:dj-def}
  \text{Pr} (d_{j} | c_{i}) =
  \begin{cases}
    1 & \text{ if } i = j \\
    \frac{2}{11} & \text{ otherwise}.
  \end{cases}
\end{align}

All that remains is to estimate $\text{Pr}(w_{jp} | d_{jp}, c_{i})$. For
this task, we borrow from the cascade click model~\cite{govindaraj2014modeling} and
write the probability as a product of the following two factors:
\begin{itemize}
\item the probability that the $p$-th position is examined by an user,
  the so-called position bias. It is denoted as $b_{p}$.
\item the intrinsic relevance of the word cloud $j$ to the color palette
  $i$. This is the quantity that we seek to infer from the user
  responses; and we will denote it as $r_{ij}$.
\end{itemize}
In other words,
\begin{align}
  \label{eq:wjp}
  \text{Pr}(w_{jp} | d_{jp}, c_{i}) = r_{ij} \cdot b_{p},
\end{align}
and by using \eqref{eq:uij-def1} and letting $q_{ij} = \sum_{p \in \{1, 2, 3\}}
\text{Pr} (d_{jp} | c_{i}) \cdot b_{p}$, we can write
\begin{align}
  \label{eq:wjp-full}
  u_{ij} = r_{ij} \cdot \sum_{p \in \{1, 2, 3\}} \text{Pr} (d_{jp} |
  c_{i}) \cdot b_{p} = r_{ij} \cdot q_{ij}.
\end{align}
Note that $b_{p}$ can be pre-computed as follows:
\begin{align}
  \label{eq:bp}
  b_{p} = \frac{m_{p}}{m},
\end{align}
where $m$ denotes the total number of trials (each question in our
survey is equivalent to one trial), and $m_{p}$ denotes the number of
times the word cloud at position $p$ was selected in any of the
trials.
In our experiments, we found the position bias of the options (in the vertical order) for the first set of the questions to be $0.323, 0.3626, 0.3381, 0.1499$ (in turn); and $0.4089, 0.3884, 0.351, 0.1364$ (in turn) for the second set of the questions. Note that for each set, these numbers do not sum up to $1$, because of the fact that the participant could choose more than one word cloud.

These numbers indicate that the position bias for each option is not equal, and even though we shuffled the three choices of word clouds in the first three vertical positions, we need to account for the position bias. We note that the fourth option ---``None of the above''--- is clicked less than the other options. This indicates that our participants wished to provide an answer, as well as the fact that they may have not thought that the associations between the word clouds and the colors were too abstract. We also note that in the second set of the questions, the fourth number is lower than the one in the first set of questions. This perhaps means that the participants who chose to participate in one more set of the questions in the survey were more confident regarding their conclusions.

As the last step, let $m_{i}$ denote the number of trials in which the $i$-th color
palette was displayed, and $m_{ij}$ denote the number of trials in which
the $i$-th color palette was displayed and the $j$-th word cloud was
selected. We can assume that the trials are independent, and therefore
the probability of observing this data under model \eqref{eq:wjp-full}
can be written as
\begin{align}
  \label{eq:multinomial}
  \text{Pr}(m_{i}, m_{ij}) = (r_{ij} \cdot q_{ij})^{m_{ij}} (1
  - r_{ij} \cdot q_{ij})^{m_{i} - m_{ij}}.
\end{align}
The maximum likelihood estimate for $r_{ij} \cdot q_{ij}$ is simply
$\frac{m_{ij}}{m_{i}}$, from which we can infer $\hat{r}_{ij}$, the
maximum likelihood estimate for $r_{ij}$, as
\begin{align}
  \label{eq:rij-mle}
  \hat{r}_{ij} = \frac{m_{ij}}{m_{i} \cdot q_{ij}}.
\end{align}

\begin{figure}[h!tb]
  \centering
  \subfloat[][]{ \includegraphics[scale=.49]{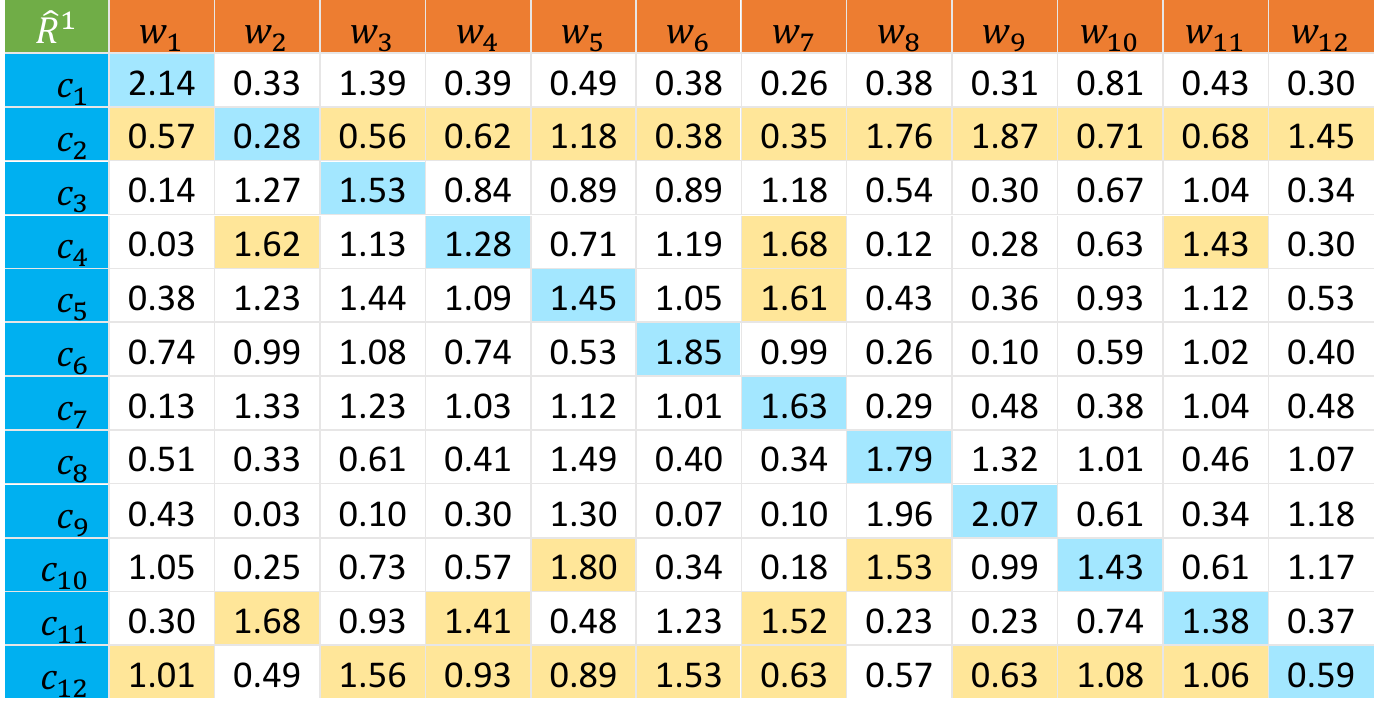}}
  \subfloat[][]{ \includegraphics[scale=.49]{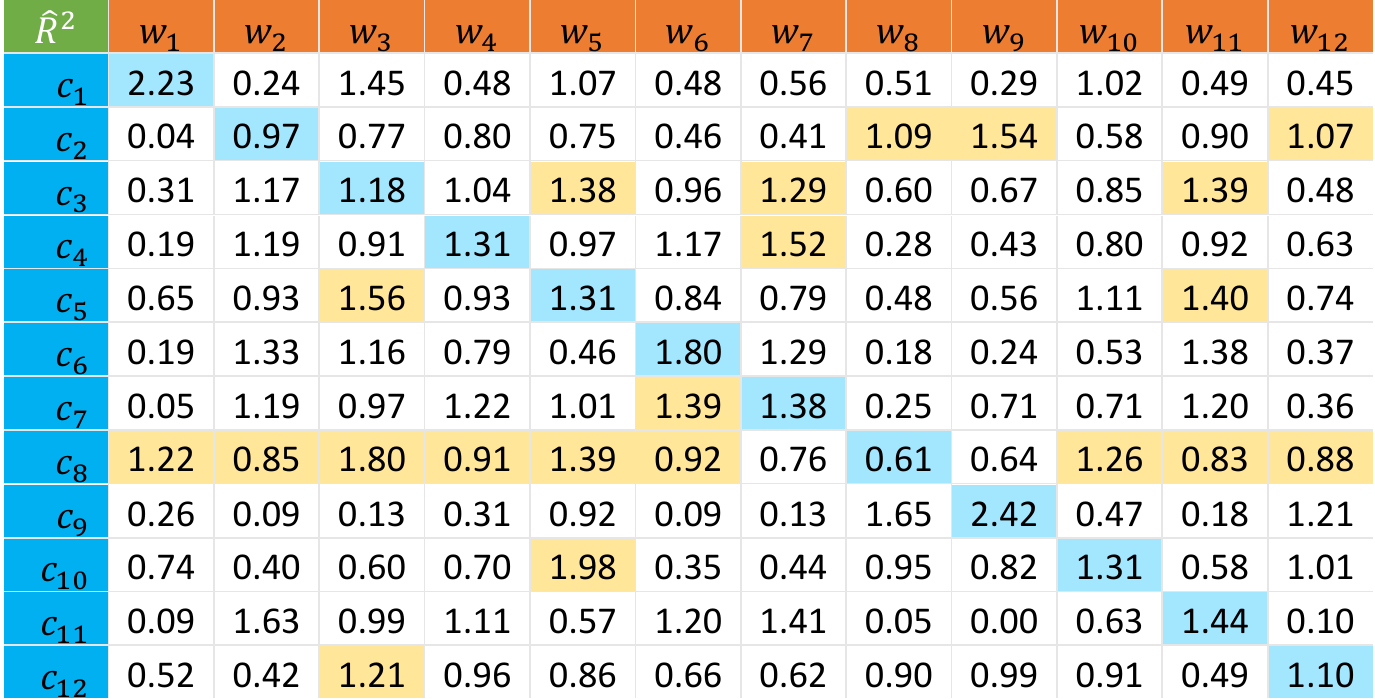}}
  \caption{Relevance matrices $\hat{R}^1$ and $\hat{R}^2$ for the first and second set
    of questions, respectively. For the first set of questions, the participants were shown the closest palettes identified by LDA-dual. For the second set of questions, the subjects were shown the second closest palettes identified by LDA-dual. The elements of these matrices are the
    estimated intrinsic relevance of associations between colors and
    words, calculated from the participants' responses. The higher the
    value, the greater the intrinsic relevance associated by the users.
    Ideally, the diagonals should contain the highest values. }
  \label{confusionMatrices}
\end{figure}

\subsection{Analyzing the Results}

Figure~\ref{confusionMatrices} illustrates two relevance matrices
$\hat{R}^{1}$ and $\hat{R}^{2}$. $\hat{R}^{1}$ (resp.\ $\hat{R}^{2}$) corresponds to the
inferred relevance of the first (resp.\ second) closest color palette to
the word cloud produced by LDA-dual. The rows are color palettes, as
proxies to color topics histograms, and the columns are word clouds, as
proxies to word topics histograms.  The $(i,j)$-th elements of these
matrices are the intrinsic relevance values $\hat{r}_{ij}$, computed
from the observed responses of the participants using the model
described in the previous section. Higher values of $\hat{r}_{ij}$ mean that the users
found a high correlation between the $i$-th word cloud and the $j$-th
color palette. If the participants find the word cloud produced by our
model to be the most relevant for a given color palette, then the
diagonal entries, marked in blue, should contain the highest values.
Whenever an off-diagonal entry is larger than the corresponding
diagonal entry, it is marked yellow in the figure.

Note that for the first set of 12 color palettes, subjects selected the ``None of the above'' option on average for $13.42\%$ of the trials. The minimum was $7\%$, which occurred for $k_1$; and the maximum was $28\%$, which occurred for $k_2$. For the second set of 12 color palettes, the average was $12.83\%$. The minimum was $2\%$, which occurred for $k_1$; and the maximum was $22\%$, which occurred for $k_2$. Frequent selection of the ``None of the above'' option for a given color palette suggests that participants had more difficulty associating this palette with the word clouds that were shown. In the relevance matrices, we do not compute the numbers for each color palette against the ``None of the above'' option. Thus the matrices do not contain the 13-th column.
However, if the participant has selected ``None of the above'' as well as other options, we take into account those options. One reason for this decision is when the participant chooses an option along with this option, he/she may feel that the other option applies but he/she is not certain about it.

\begin{figure}[H]\tiny
  \centering
  \subfloat[][]{ \includegraphics[scale=.37]{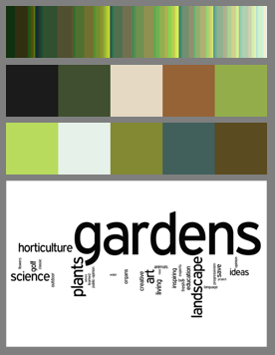}}
  \subfloat[][]{ \includegraphics[scale=.37]{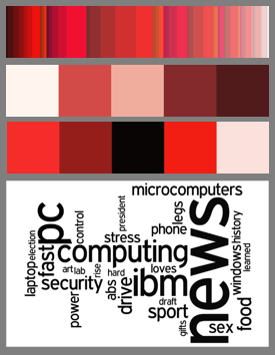}}
  \subfloat[][]{ \includegraphics[scale=.37]{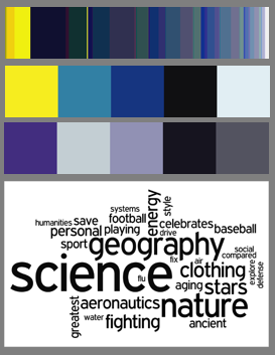}} \\
  \subfloat[][]{ \includegraphics[scale=.37]{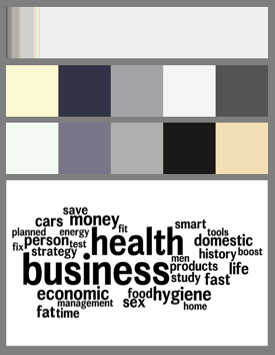}}
  \subfloat[][]{ \includegraphics[scale=.37]{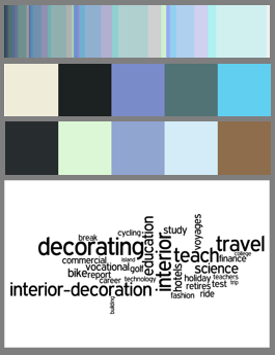}}
  \subfloat[][]{ \includegraphics[scale=.37]{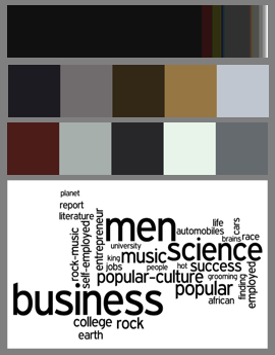}} \\
  \subfloat[][]{ \includegraphics[scale=.37]{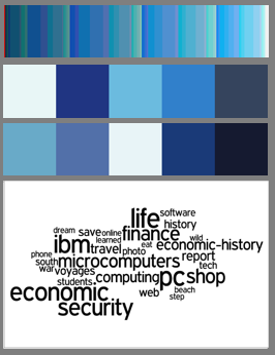}}
  \subfloat[][]{ \includegraphics[scale=.37]{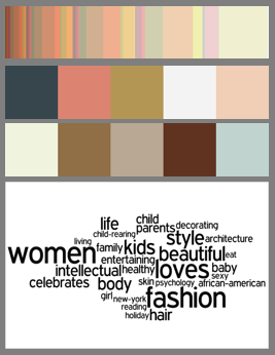}}
  \subfloat[][]{ \includegraphics[scale=.37]{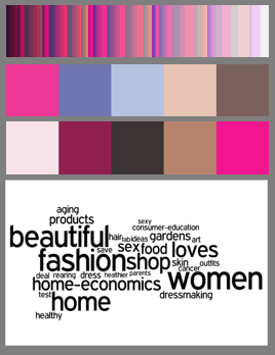}}  \\
  \subfloat[][]{ \includegraphics[scale=.37]{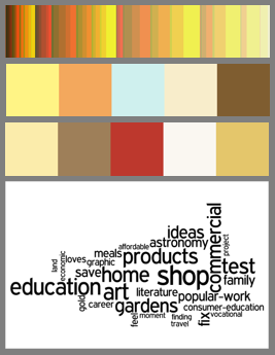}}
  \subfloat[][]{ \includegraphics[scale=.37]{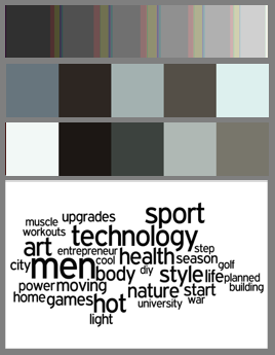}}
  \subfloat[][]{ \includegraphics[scale=.37]{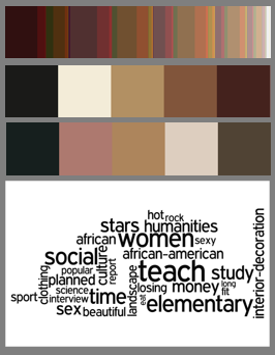}}
  \caption{The 12 color-word topics produced by the model are
    visualized in this figure. The top panel shows the color histogram,
    the second and third color panels show the top two color palettes we
    extracted from this histogram. The word topics are visualized in the
    bottom panel as word clouds, with the size of a word being
    proportional to its weight. }
  \label{color-word-topics}
\end{figure}

As it is observed from the relevance matrices in
Fig.~\ref{confusionMatrices}, in most cases the diagonal elements are
higher than the off-diagonal elements. This indicates a strong
correlation between the results of LDA-dual and the opinion of the
participants (also see the next section for an aggregated measure).  We studied the entries with the highest relevance values
such as $\hat{r}^{1}_{11}$ and $\hat{r}^{1}_{99}$.  Referring to
Figs~\ref{color-word-topics} (a) and (i), one reason these values are
high could be because green for ``garden'' ($\hat{r}^{1}_{11}$) and pink and
purple for ``fashion'' and ``women'' ($\hat{r}^{2}_{99}$) are intuitive and
widely accepted associations across different cultures.

There are a few color palettes such as $c_{2}$ in $\hat{R}^{1}$ where the
users assign higher relevance to word clouds other than the one produced
by the LDA-dual model. To understand this, note that in
Fig.~\ref{color-word-topics} (b), the first 5-color palette which is
$c_2$ predominantly contains shades of red and black. Unsurprisingly,
users assign higher relevance to word clouds $w_8$ and $w_9$, which are
about ``sex'' and ``beauty.''  In our dataset however, the red and
black color combinations are often used by news magazines such as \emph{Time}
and \emph{The Economist}, and computer magazines such as \emph{PC Magazine}. One can
perform a similar analysis for other color palettes such as $c_{12}$ in
$\hat{R}^{1}$ and $c_{8}$ in $\hat{R}^{2}$ to infer why there is a mismatch between
the model output and the relevance values assigned by the users. This
also shows why it is important to perform a user study; domain specific
color palettes and their associated linguistic concepts may not always
transfer to a general context.

\begin{figure}[H]\tiny
  \centering
  \subfloat[][]{ \includegraphics[scale=.25]{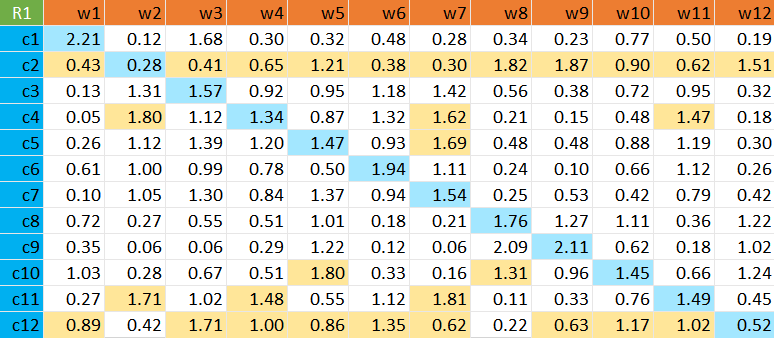}} \ \
  \subfloat[][]{ \includegraphics[scale=.25]{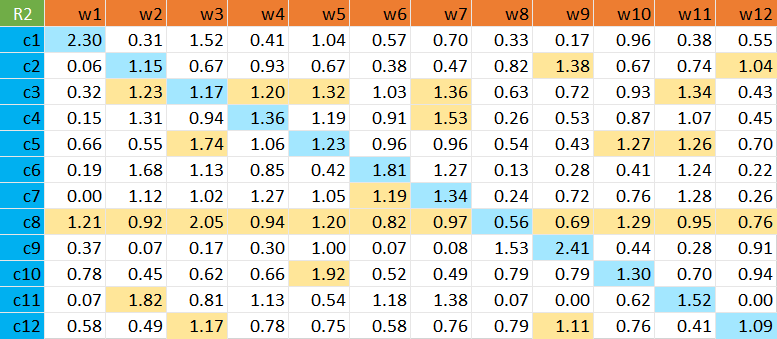}} \\
  \subfloat[][]{ \includegraphics[scale=.25]{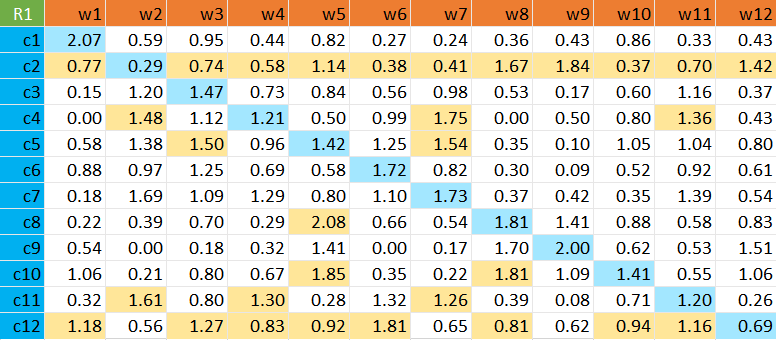}} \ \
  \subfloat[][]{ \includegraphics[scale=.25]{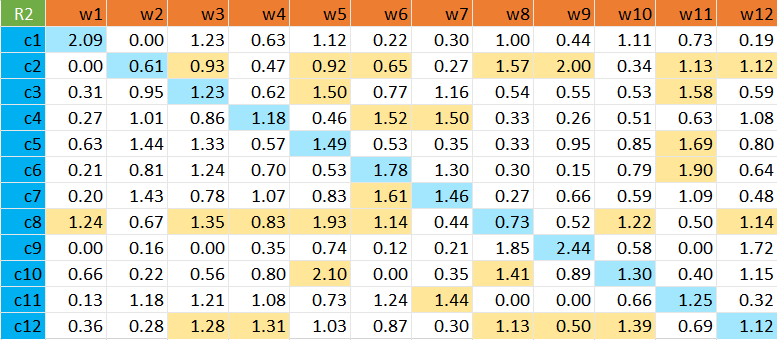}}\\
  \subfloat[][]{ \includegraphics[scale=.25]{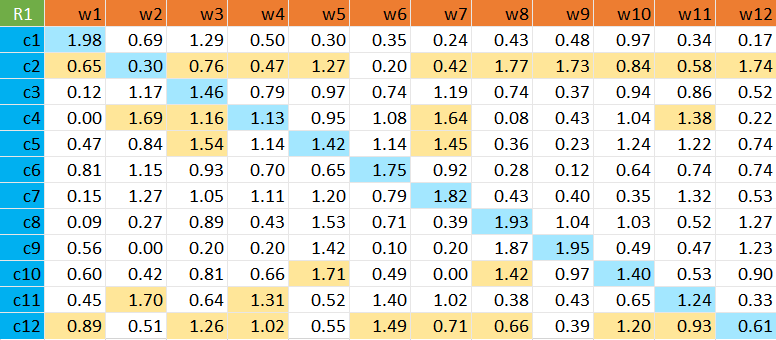}} \ \
  \subfloat[][]{ \includegraphics[scale=.25]{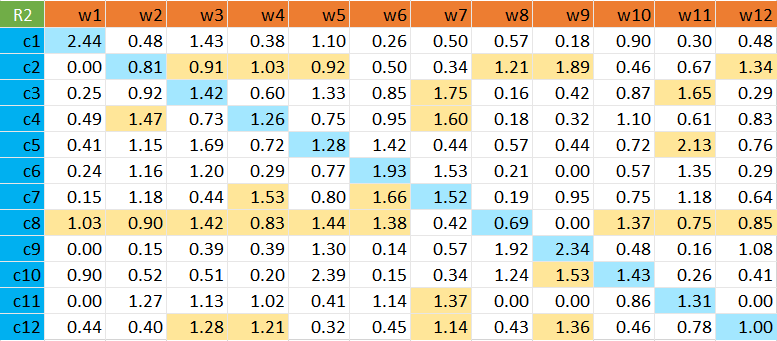}}\\
  \subfloat[][]{ \includegraphics[scale=.25]{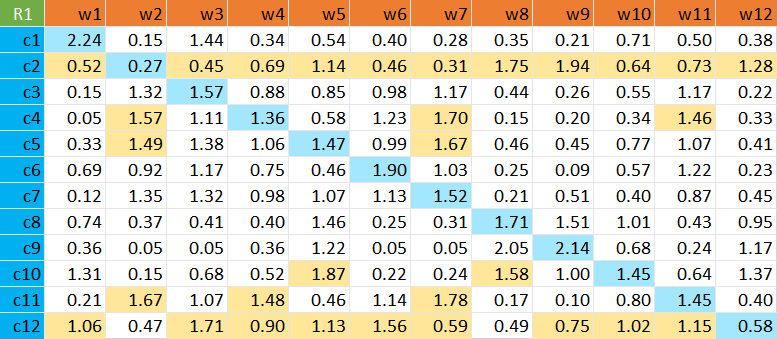}} \ \
  \subfloat[][]{ \includegraphics[scale=.25]{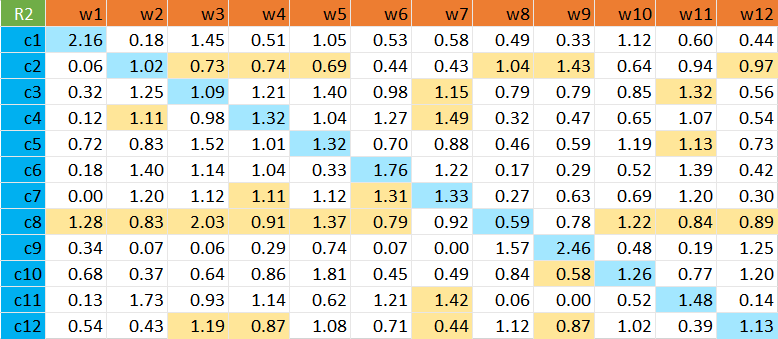}}\\
  \subfloat[][]{ \includegraphics[scale=.25]{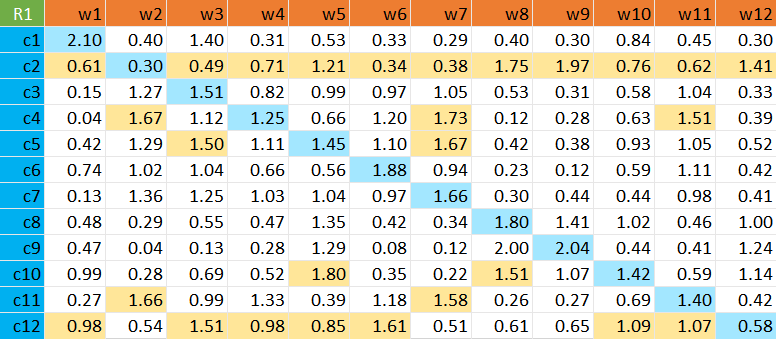}} \ \
  \subfloat[][]{ \includegraphics[scale=.25]{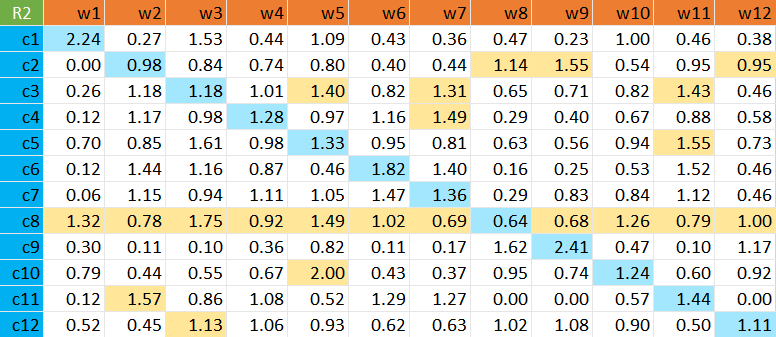}}\\
  \subfloat[][]{ \includegraphics[scale=.25]{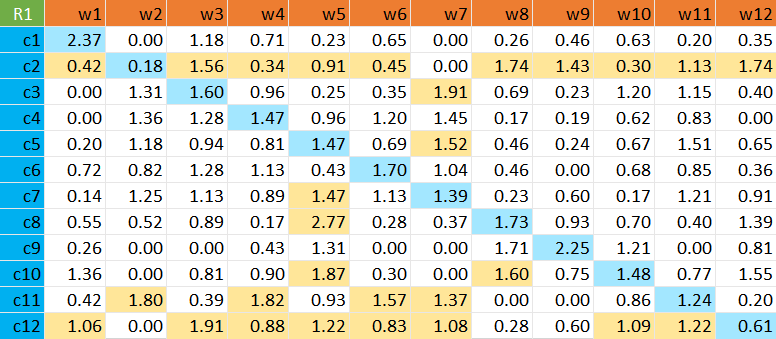}} \ \
  \subfloat[][]{ \includegraphics[scale=.25]{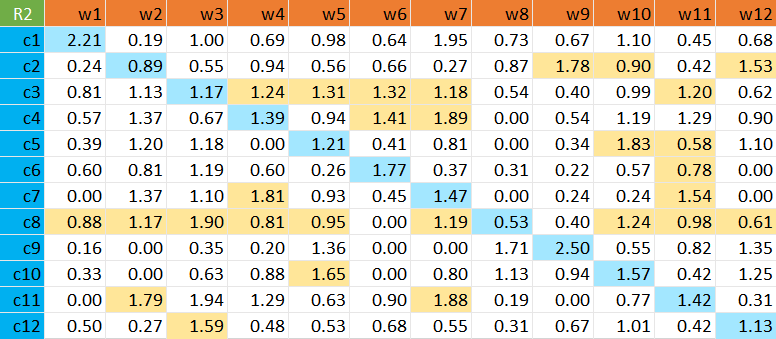}}
  \caption{Relevance matrices $\hat{R}^1$ and $\hat{R}^2$ for the first and second set
    of questions, respectively, computed for: (a) and (b) females, (c) and (d) males,
    (e) and (f) Non-US participants, (g) and (h) US participants, (i) and (j) Non-designers, and (k) and (l) designers.}
  \label{confMat_gender_vcd_diff}
\end{figure}

To understand differences between female versus male and non-US versus US
participants, we computed the corresponding relevance matrices (see Fig.~\ref{confMat_gender_vcd_diff}).
Comparing these matrices with Fig.~\ref{confusionMatrices}, we do not observe any
significant differences. This indicates that our results do not depend
significantly on the gender or cultural background of the
users. However, when we compare the designers versus non-designers, we note that there
are more zero values in the off-diagonals for designers. This indicates
that designers exhibit a stronger bias against selecting certain word
clouds for certain color palettes, perhaps because of their training~\cite{whitfield1982design}.

\subsection{Aggregated Measures}

We have defined two different measures to assess the strength of the relationships of elements in a matrix as indicated by the magnitude of the diagonal elements relative to the off-diagonal elements in each row. For the definitions below, it is assumed that the matrix elements are all non-negative, and are defined from
\begin{align}
  \label{eq:aggregatedMatrix}
  R = \left [ r_{ij} \right ],
\end{align}
for each relevance matrix $\hat{R}^{1}$ and $\hat{R}^{2}$.

The first measure is the \emph{diagonal dominance} $D$. For each row, it is simply the ratio of the diagonal element in that row to the average value of the non-diagonal elements in that row:
\begin{align}
  \label{eq:diagonalDominance}
  D_i = \frac{r_{ii}}{\bar{\acute{r_{i}}}},
\end{align}
where $r_{ii}$ is a diagonal element, and $\bar{\acute{r_{i}}}$ is the mean of off-diagonal elements:
\begin{align}
  \label{eq:meanOff-diagonal}
  \bar{\acute{r_{i}}} = \frac{1}{N-1}\sum_{\substack{j = 1 \\ j \neq i}}^{N} r_{ij}.
\end{align}

\begin{figure}[h!tb]
  \centering
    \includegraphics[scale=1]{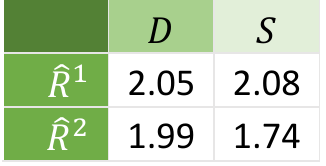}
  \caption{Aggregated measures for relevance matrices $\hat{R}^1$ and $\hat{R}^2$. Column $D$ is the diagonal dominance measure computed according to (\ref{eq:averageDiagonalDominance}). Column $S$ is the diagonal separation computed according to (\ref{eq:diagonalSeparation}). Overall, these two separate measures support the conclusion that on average, the diagonal elements in the relevance matrices are significantly stronger than the off-diagonal elements.}
  \label{confusionMatrices_aggregate}
\end{figure}

To get a summary assessment, we can average the diagonal dominance over all the rows of the matrix:
\begin{align}
  \label{eq:averageDiagonalDominance}
  \bar{D} = \frac{1}{N} \sum_{i = 1}^{N} \bar{\acute{r_{i}}}.
\end{align}
For the two matrices $\hat{R}^{1}$ and $\hat{R}^{2}$ in Fig.\ref{confusionMatrices}, the average diagonal dominances are $2.05$ and $1.99$, respectively (see Fig.~\ref{confusionMatrices_aggregate}). So on average, the diagonal element is twice as large as the other elements in the row. This suggests that despite the wide variability of the data in the matrices, the diagonal elements do tend to dominate.

The second measurement is the \emph{diagonal separation} $S$. It is also defined for each row, and is also a ratio:
\begin{align}
  \label{eq:diagonalSeparation}
  S_i = \frac{\left| r_{ii} - \bar{\acute{r_{i}}} \right|}{_r\acute{\sigma}_i}.
\end{align}
The numerator is the absolute value of the difference between the diagonal element and the mean of the off-diagonal elements in that row. The denominator is the standard deviation of the off-diagonal elements in that row:
\begin{align}
  \label{eq:std_diagonalSeparation}
  _r\acute{\sigma}_i = \left( \frac{1}{N-1} \sum_{\substack{j = 1 \\ j \neq i}}^{N} \left(r_{ij} - \bar{\acute{r_{i}}} \right)^2  \right)^{\frac{1}{2}}.
\end{align}

To get a summary assessment, we can again average the diagonal separation over all the rows of the matrix:
\begin{align}
  \label{eq:diagonalSeparation}
  \bar S = \frac{1}{N} \sum_{i = 1}^{N} {_r\acute{\sigma}_i}.
\end{align}

As Fig.~\ref{confusionMatrices_aggregate} summarizes, for matrix $\hat{R}^{1}$ (Fig.~\ref{confusionMatrices} (a)), the average diagonal separation is $2.08$. Thus, on average, the diagonal element is more than two standard deviations away from the mean of the off-diagonal elements. This is a very good level of separation. For the second matrix, $\hat{R}^{1}$ (Fig.~\ref{confusionMatrices} (b)), the average diagonal separation is $1.74$. While this is not as large as it is for the first matrix , it still indicates very good separation.

Overall, these two separate measures support the conclusion that on average, the diagonal elements are, indeed, significantly stronger than the off-diagonal elements.

\section{Applications}
\label{sec:Applications}

Similar to the applications of color naming proposed by Heer and Stone~\cite{heer2012color},
in order to illustrate how color semantics enables more meaningful user interactions, we present a number of applications in color palette selection and design example retrieval, image retrieval, and recoloring images using semantics.

\subsection{Color Palettes Selection Using Semantics}

One way to make design accessible to the masses is to recommend design
elements and design examples.  In content creation, for instance, the
user needs to use a color combination that is both appealing and aligned
with his/her purpose of design.  While this need seems more immediate for the non-designer, a designer may also prefer to see examples for a
more creative or even an inspiring color combination.  Our inferred
color semantics can be applied in color palette recommendation.
Although there are pools of color palettes available
online~\cite{AdobeKuler,colourlovers}, it is not easy to navigate, find,
or recommend a set of palettes based on the user's need (see~\cite{o2011color} and~\cite{O'Donovan:2014:CFC:2630099.2630100}).
Consider a scenario in which the user wishes to find a color palette that conveys meanings of both ``technology'' and ``fashion.''
Figure~\ref{teaser_img} illustrates such a scenario, where the user queries for ``techy fashion''. This type of query is mapped to the word topics, and from them to their corresponding color topics. We then map these color topics to a set of color palettes in a pool of color palettes, created from the magazine dataset using techniques in~\cite{2013-color-themes}. The user can then choose his/her preferred color palettes from the recommended set.

\subsection{Design Example Recommendation}

Another plausible application that could use our color semantics is
recommendation of alternative designs. This can be done in several ways, i.e. by using a given a set of recommended color palettes (see Fig.~\ref{teaser_img}), or directly, just by taking color semantics as the user's input. 
Providing such examples can be utilized in creativity
support tools~\cite{shneiderman2009creativity}.  It can also facilitate
design prototyping~\cite{dow2010parallel}.  Similarly, translating the
associations inferred from the magazine covers to other kinds of media
such as website remains a topic for future work.  Because LDA-dual is a generative
model, it can be used in automatic creation of designs based on the
user's preferences, similar to magazine cover design~\cite{jahanian2013recommendation},
and website design~\cite{kumar2011bricolage}.  Utilizing generative
models can also leverage the design-driven studies in
HCI~\cite{talton2012learning}.

\subsection{Image Retrieval Using Color Semantics}

\begin{figure}[h!tb]
  \centering
  \includegraphics[width=.8\textwidth]{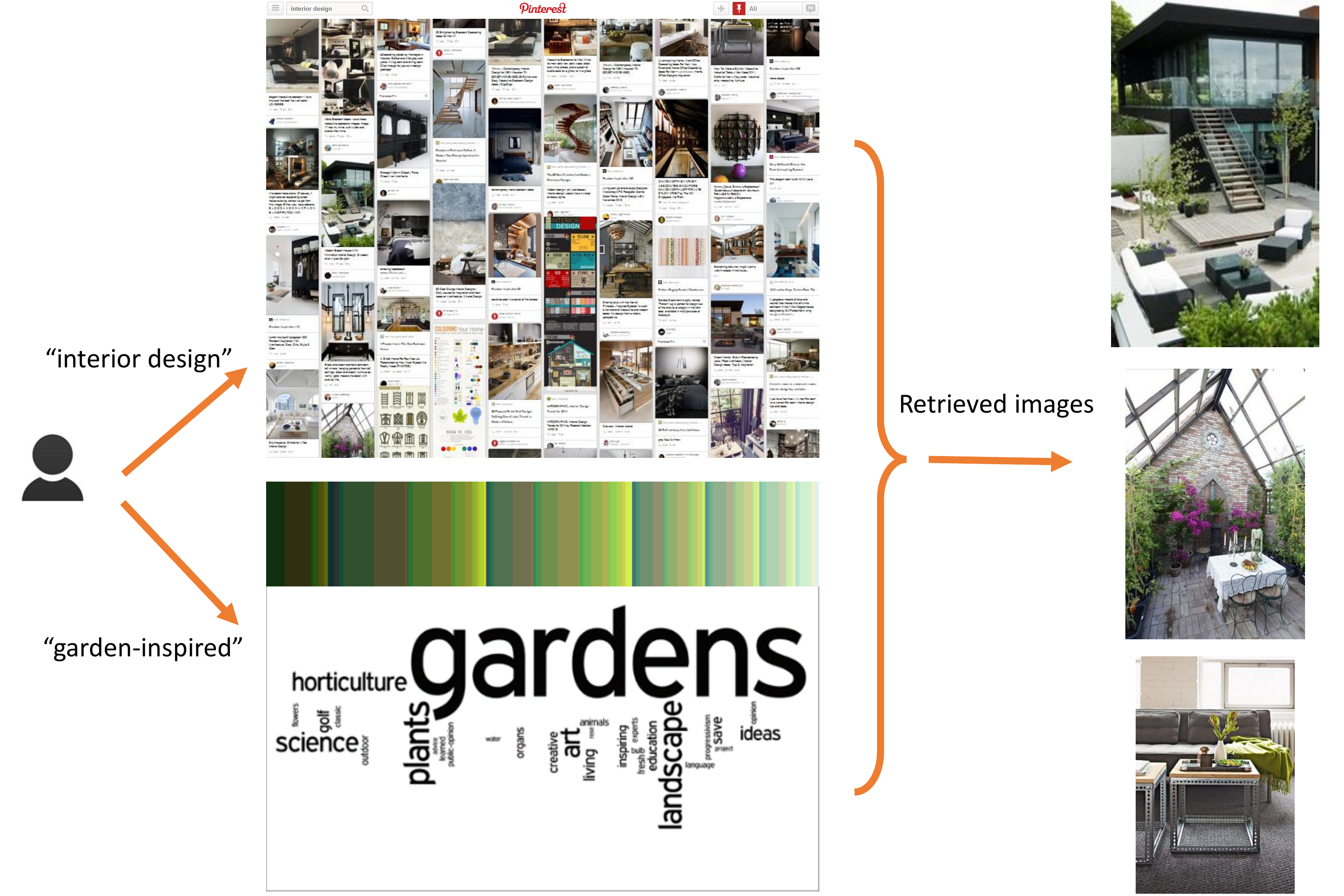}
  \caption{Hypothetical application of image retrieval using semantics. The retrieved images on the right side are ranked vertically. In this vertical order, a larger image represents a higher rank. Images from \protect\cite{pinterest}.} \label{img_retrieval_app}
\end{figure}

The gap of color semantics in computer vision,
specifically for image retrieval, has been emphasized by prior work~\cite{smeulders2000content,sethi2001mining,mojsilovic2001capturing,liu2007survey}.
Nevertheless, the associations of color combinations and linguistic
concepts can be utilized in designing high level image features for both
current image retrieval algorithms~\cite{solli2010color} as well as deep learning algorithms~\cite{couprie2013indoor,socher2013grounded}.
Figure~\ref{img_retrieval_app} illustrates an application of our inferred color semantics in image retrieval. Consider a scenario in which, the user first makes a query in an image search community, e.g. Pinterest~\cite{pinterest} about ``interior design''. As a results, a number of images will be retrieved and suggested to the user. However, in order to explore and navigate through the retrieved images, the user may query for ``garden-inspired'' images. In this case, we are able to map this query to the color-word topics, and rank the already retrieved images based on the color topic histograms that represent ``garden''.

\subsection{Image Color Selection Using Semantics}

\begin{figure}[h!tb]\tiny
  \centering
  \subfloat[][]{ \includegraphics[scale=.9]{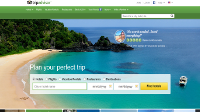}} \ \
  \subfloat[][]{ \includegraphics[scale=.9]{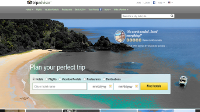}} \ \
  \caption{Image color selection using color semantics. (a) The original image, (b) colors that contribute to ``travel'' and ``shop'' in the original image. Image from the home page of Tripadvisor.com~\protect\cite{tripadvisor}.}
  \label{color_selection}
\end{figure}

As Heer and Stone~\cite{heer2010crowdsourcing} note, a common interaction in image editing is to find a region of colors. They suggest two types of color region mechanisms based on color names: color name queries from the user, and using the \emph{magic wand} tool.
In our case, we use the user's query and present the set of pixels that semantically contributes to the colors of an image.
Figure~\ref{color_selection} illustrates this kind of interaction. Figure~\ref{color_selection} (a) is the original image, a screenshot of a travel agency website~\cite{tripadvisor} nominated as one of the best designs (based on people's votes) according to the Webby Awards 2014~\cite{webbyawards}.
We are interested to know what color regions have contributed to ``travel'' and ``shop'' in this image.
Figure~\ref{color_selection} (b) represents these regions, while turning the other regions to grayscale mode.

Note that in order to find these color regions we map the user's queries to the word topics, and preserve their associated color topics in the image. Of course, utilizing the current image segmentation techniques leads to a smoother pixel selection.

\subsection{Image Recoloring Using Semantics}

\begin{figure}[h!tb]\tiny
  \centering
  \subfloat[][]{ \includegraphics[scale=.4]{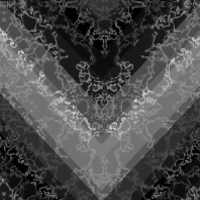}} \ \
  \subfloat[][]{ \includegraphics[scale=.4]{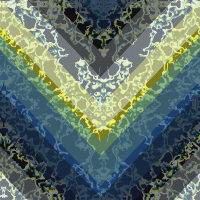}} \ \
  \subfloat[][]{ \includegraphics[scale=.4]{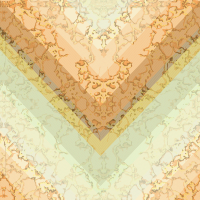}} \ \
  \subfloat[][]{ \includegraphics[scale=.4]{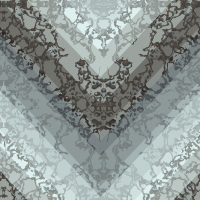}}
  \caption{Pattern colorization using color semantics. (a) The original pattern, (b) ``science'', (c) ``shop'', and (d) ``sports'' and ``men'' colorized patterns. Images colorized using techniques in~\protect\cite{2013-patternColoring}. Pattern from \emph{ColourLovers ArrayOfLilly}~\protect\cite{colourlovers}.}
  \label{pattern_colorization}
\end{figure}

Given the fact that colors communicate messages, designers often modify color themes of their designs. Part of this color modification may involve recoloring an image~\cite{2013-patternColoring}, transferring a color theme to an image~\cite{murray2012toward}, or enhancing a color theme of an image~\cite{wang2010data}.
In this section we suggest an application of color semantics for colorizing patterns (Fig.~\ref{pattern_colorization}), based on techniques introduced by Lin et al.~\cite{2013-patternColoring}\footnote{The recolored patterns in this paper are produced by S. Lin, at our request.}.

Figure~\ref{pattern_colorization} (a) illustrates a grayscale pattern.
Using the color semantics associations, we are able to accept a user query, map it to the word topics, and use a 5-color palette as a representative of this color topic to colorize the original pattern.
Figures~\ref{pattern_colorization} (b), (c), and (d) illustrate the results of colorizing the original pattern using ``science'', ``shop'', and ``sports'' and ``men'' queries, respectively.

\section{Conclusion and Future Work}
\label{sec:ConclusionAndFutureWork}

Taking into account that the goal of visual design is both to convey a
message and to appeal to audiences aesthetically, we investigate design
mining of designers' thought processes in associating colors with
linguistic concepts.  We collected high quality examples of good
designs, a dataset of 2,654 magazine cover designs in 71 titles and 12
genres.  We then adapted LDA-dual, an extension of the popular LDA topic
model, to simultaneously explain both designers' choice of colors
and choice of words on the magazine covers.  To verify the inferred color
semantics with the wisdom of the crowd, we used a crowdsourcing experiment.
The results confirm that our model is able to successfully discover the
association between colors and linguistic concepts.  This completes the
loop of our design mining system, from hypothesis inference to
validation.

Our user study shows that users across different countries, gender, and
age groups largely agree with the colors and linguistic concepts
associations produced by our model. However, to put this in context, note
that our experiment is biased toward an audience that is capable of
reading and understanding the English language, is mainly college educated, and has
access to the Internet. Arguably, the Internet is globalizing design by
melting cultural perceptions~\cite{carroll2007art}.
Nevertheless, studying color semantics in different cultures and
folklore (similar to~\cite{reinecke2014} in studying aesthetics of low level features of colors) can increase our understanding of customized designs and
hopefully help both designers and non-designers in communicating with
their audiences.

We then suggest a number of applications for color semantics to demonstrate how semantics can enable more meaningful user interactions, and perhaps help masses to design better.
We specifically discuss color palette selection and design example recommendation, image retrieval, color region selection in images, and pattern colonization in image recoloring.
Nevertheless, at this stage, these applications are not final tools. They need to be investigated more in terms of user experience design, and to be evaluated how much they can actually facilitate user interactions. Incorporating these applications in current design creation tools is also part of the future work.

The present work is a first step towards our broader
goal of making design accessible to the general public.  We believe that this
work can open up a number of interesting avenues for future work, some
of which we now discuss.

\subsection{Visual Design Language and User Interaction}

From a broader perspective, association of linguistic concepts with
design elements, specifically colors, can accelerate the development of
a visual design language.  This can enhance the user interaction
mechanisms to be more intuitive and cognitive, and the user's
involvement in design tasks such as prototyping, collaboration, and
critique.  The LDA-dual model can be utilized to associate the critiques
(in a design critique system such as~\cite{xu2014voyant}) and semantic
tags (in a design query system such as~\cite{kumar2013webzeitgeist})
with design elements of the corresponding designs.  This may provide
more insight to understanding design critiques and queries.

\subsection{Quantifying Aesthetics}

Current computational models for quantifying aesthetics are based on low
level features of design
elements~\cite{datta2006studying,li2009aesthetic,joshi2011aesthetics,reinecke2013predicting}.
We believe that both outcomes of this work, the inferred color semantics and
the application of LDA-dual model to color semantics, can leverage the current models of quantifying
aesthetics. While the first outcome is intended to provide more high
level features to evaluate a design based on the message that it is
meant to convey, the second outcome can be used in linking linguistic
concepts with colors, spatial composition (layout), and typography of visual
media design.

\newpage
\section*{APPENDIX}
\setcounter{section}{0}
\label{app:Data}
\scriptsize\label{table:magazine_cover_dataset}
\begin{longtable}[!htb]{|l|l|l|l|}
    \caption{Summary of Our Magazine Covers Dataset.}\\
   \hline
  \endfirsthead
    \caption[]{\emph{continued}}\\
    \hline
  \endhead
    \hline
    \multicolumn{4}{r}{\emph{continued on next page}}
  \endfoot
    \hline
  \endlastfoot
\rowcolor[HTML]{F56B00}
\multicolumn{2}{|c|}{\cellcolor[HTML]{F56B00}{\color[HTML]{1F497D} \textbf{Art}}}                           & \multicolumn{2}{c|}{\cellcolor[HTML]{F56B00}{\color[HTML]{1F497D} \textbf{Business}}}          \\ \hline
\rowcolor[HTML]{C0C0C0}
\textbf{Magazine Title}                                      & \textbf{\# Collected}                        & \textbf{Magazine Title}                         & \textbf{\# Collected}                        \\ \hline
\cellcolor[HTML]{EFEFEF}Art in America                       & 49                                           & \cellcolor[HTML]{EFEFEF}Bloomberg Businessweek  & 50                                           \\ \hline
\cellcolor[HTML]{EFEFEF}ARTNews                              & 50                                           & \cellcolor[HTML]{EFEFEF}Entrepreneur            & 52                                           \\ \hline
\cellcolor[HTML]{EFEFEF}HOW                                  & 41                                           & \cellcolor[HTML]{EFEFEF}Forbes                  & 50                                           \\ \hline
\cellcolor[HTML]{EFEFEF}ID                                   & 1                                            & \cellcolor[HTML]{EFEFEF}Harvard Business Review & 50                                           \\ \hline
\cellcolor[HTML]{EFEFEF}Interior Design                      & 50                                           & \cellcolor[HTML]{EFEFEF}Money                   & 48                                           \\ \hline
\cellcolor[HTML]{EFEFEF}Live Design                          & 9                                            & \cellcolor[HTML]{EFEFEF}The Economist           & 29                                           \\ \hline
\cellcolor[HTML]{EFEFEF}The New Yorker                       & 50                                           &                                                 &                                              \\ \hline
                                                             &                                              &                                                 &                                              \\ \hline
\rowcolor[HTML]{EFEFEF}
Art Total:                                                   & 250                                          & Business Total:                                 & 279                                          \\ \hline
\rowcolor[HTML]{F56B00}
\multicolumn{2}{|c|}{\cellcolor[HTML]{F56B00}{\color[HTML]{1F497D} \textbf{Education}}}                     & \multicolumn{2}{c|}{\cellcolor[HTML]{F56B00}{\color[HTML]{1F497D} \textbf{Entertainment}}}     \\ \hline
\rowcolor[HTML]{C0C0C0}
\textbf{Magazine Title}                                      & \textbf{\# Collected}                        & \textbf{Magazine Title}                         & \textbf{\# Collected}                        \\ \hline
\cellcolor[HTML]{EFEFEF}Academe                              & 40                                           & \cellcolor[HTML]{EFEFEF}Conde Nast Traveler     & 50                                           \\ \hline
\cellcolor[HTML]{EFEFEF}Language                             & 36                                           & \cellcolor[HTML]{EFEFEF}Jet                     & 50                                           \\ \hline
\cellcolor[HTML]{EFEFEF}Language Arts                        & 2                                            & \cellcolor[HTML]{EFEFEF}National Geographic     & 44                                           \\ \hline
\cellcolor[HTML]{EFEFEF}Social Studies and the Young Learner & 50                                           & \cellcolor[HTML]{EFEFEF}People                  & 50                                           \\ \hline
\cellcolor[HTML]{EFEFEF}Teaching Exceptional Children        & 2                                            & \cellcolor[HTML]{EFEFEF}Rolling Stone           & 50                                           \\ \hline
\cellcolor[HTML]{EFEFEF}Techniques                           & 48                                           &                                                 &                                              \\ \hline
\cellcolor[HTML]{EFEFEF}The American Scholar                 & 3                                            &                                                 &                                              \\ \hline
\cellcolor[HTML]{EFEFEF}The Reading Teacher                  & 3                                            &                                                 &                                              \\ \hline
\cellcolor[HTML]{EFEFEF}The Science Teacher                  & 50                                           &                                                 &                                              \\ \hline
                                                             &                                              &                                                 &                                              \\ \hline
\rowcolor[HTML]{EFEFEF}
Education Total:                                             & 234                                          & Entertainment Total:                            & 244                                          \\ \hline
\rowcolor[HTML]{F56B00}
\multicolumn{2}{|c|}{\cellcolor[HTML]{F56B00}{\color[HTML]{1F497D} \textbf{Family}}}                        & \multicolumn{2}{c|}{\cellcolor[HTML]{F56B00}{\color[HTML]{1F497D} \textbf{Fashion}}}           \\ \hline
\rowcolor[HTML]{C0C0C0}
\textbf{Magazine Title}                                      & \textbf{\# Collected}                        & \textbf{Magazine Title}                         & \textbf{\# Collected}                        \\ \hline
\cellcolor[HTML]{EFEFEF}Good Housekeeping                    & 61                                           & \cellcolor[HTML]{EFEFEF}Esquire                 & 11                                           \\ \hline
\cellcolor[HTML]{EFEFEF}Better Homes and Gardens             & 30                                           & \cellcolor[HTML]{EFEFEF}Essence                 & 50                                           \\ \hline
\cellcolor[HTML]{EFEFEF}House Beautiful                      & 50                                           & \cellcolor[HTML]{EFEFEF}Glamour                 & 50                                           \\ \hline
\cellcolor[HTML]{EFEFEF}Parenting                            & 51                                           & \cellcolor[HTML]{EFEFEF}GQ                      & 50                                           \\ \hline
\cellcolor[HTML]{EFEFEF}ShopSmart                            & 40                                           & \cellcolor[HTML]{EFEFEF}Vogue                   & 50                                           \\ \hline
                                                             &                                              &                                                 &                                              \\ \hline
\rowcolor[HTML]{EFEFEF}
Family Total:                                                & 232                                          & Fashion Total:                                  & 211                                          \\ \hline
\rowcolor[HTML]{F56B00}
\multicolumn{2}{|c|}{\cellcolor[HTML]{F56B00}{\color[HTML]{1F497D}\textbf{Health}}}                                               & \multicolumn{2}{c|}{\cellcolor[HTML]{F56B00}{\color[HTML]{1F497D}\textbf{Nature}}}                                   \\ \hline
\rowcolor[HTML]{C0C0C0}
{\color[HTML]{333333} \textbf{Magazine Title}}               & {\color[HTML]{333333} \textbf{\# Collected}} & {\color[HTML]{333333} \textbf{Magazine Title}}  & {\color[HTML]{333333} \textbf{\# Collected}} \\ \hline
\cellcolor[HTML]{EFEFEF}Health                               & 17                                           & \cellcolor[HTML]{EFEFEF}BBC Wildlife            & 25                                           \\ \hline
\cellcolor[HTML]{EFEFEF}Men's Health                         & 50                                           & \cellcolor[HTML]{EFEFEF}Garden Design           & 35                                           \\ \hline
\cellcolor[HTML]{EFEFEF}Natural Health                       & 17                                           & \cellcolor[HTML]{EFEFEF}House \& Garden         & 15                                           \\ \hline
\cellcolor[HTML]{EFEFEF}Women's Health                       & 50                                           & \cellcolor[HTML]{EFEFEF}Organic Gardening       & 17                                           \\ \hline
                                                             &                                              &                                                 &                                              \\ \hline
\rowcolor[HTML]{EFEFEF}
Health Total:                                                & 134                                          & Nature Total:                                   & 92                                           \\ \hline
\rowcolor[HTML]{F56B00}
\multicolumn{2}{|c|}{\cellcolor[HTML]{F56B00}{\color[HTML]{1F497D}\textbf{Politics}}}                                             & \multicolumn{2}{c|}{\cellcolor[HTML]{F56B00}{\color[HTML]{1F497D} \textbf{Science}}}                                  \\ \hline
\rowcolor[HTML]{C0C0C0}
\textbf{Magazine Title}                                      & \textbf{\# Collected}                        & \textbf{Magazine Title}                         & \textbf{\# Collected}                        \\ \hline
\cellcolor[HTML]{EFEFEF}Human Rights                         & 42                                           & \cellcolor[HTML]{EFEFEF}Astronomy               & 10                                           \\ \hline
\cellcolor[HTML]{EFEFEF}Newsweek                             & 50                                           & \cellcolor[HTML]{EFEFEF}Nature                  & 50                                           \\ \hline
\cellcolor[HTML]{EFEFEF}The Atlantic                         & 25                                           & \cellcolor[HTML]{EFEFEF}Popular Science         & 29                                           \\ \hline
\cellcolor[HTML]{EFEFEF}The Progressive                      & 22                                           & \cellcolor[HTML]{EFEFEF}Science                 & 119                                          \\ \hline
\cellcolor[HTML]{EFEFEF}Time                                 & 50                                           & \cellcolor[HTML]{EFEFEF}Science and Children    & 5                                            \\ \hline
                                                             &                                              & \cellcolor[HTML]{EFEFEF}Science Illustrated     & 30                                           \\ \hline
                                                             &                                              & \cellcolor[HTML]{EFEFEF}Smithsonian             & 38                                           \\ \hline
                                                             &                                              &                                                 &                                              \\ \hline
\rowcolor[HTML]{EFEFEF}
Politics Total:                                              & 189                                          & Science Total:                                  & 281                                          \\ \hline
\rowcolor[HTML]{F56B00}
\multicolumn{2}{|c|}{\cellcolor[HTML]{F56B00}{\color[HTML]{1F497D}\textbf{Sports}}}                                               & \multicolumn{2}{c|}{\cellcolor[HTML]{F56B00}{\color[HTML]{1F497D}\textbf{Technology}}}                               \\ \hline
\rowcolor[HTML]{C0C0C0}
\textbf{Magazine Title}                                      & \textbf{\# Collected}                        & \textbf{Magazine Title}                         & \textbf{\# Collected}                        \\ \hline
\cellcolor[HTML]{EFEFEF}Bicycling                            & 22                                           & \cellcolor[HTML]{EFEFEF}Aviation Week           & 45                                           \\ \hline
\cellcolor[HTML]{EFEFEF}Car and Driver                       & 29                                           & \cellcolor[HTML]{EFEFEF}PC World                & 50                                           \\ \hline
\cellcolor[HTML]{EFEFEF}Golf Digest                          & 26                                           & \cellcolor[HTML]{EFEFEF}Popular Mechanics       & 34                                           \\ \hline
\cellcolor[HTML]{EFEFEF}Golf World                           & 24                                           & \cellcolor[HTML]{EFEFEF}Technology Review       & 17                                           \\ \hline
\cellcolor[HTML]{EFEFEF}Runner's World                       & 22                                           & \cellcolor[HTML]{EFEFEF}Wired                   & 45                                           \\ \hline
\cellcolor[HTML]{EFEFEF}Sporting News                        & 44                                           &                                                 &                                              \\ \hline
\cellcolor[HTML]{EFEFEF}Sports Illustrated                   & 50                                           &                                                 &                                              \\ \hline
                                                             &                                              &                                                 &                                              \\ \hline
\rowcolor[HTML]{EFEFEF}
Sports Total:                                                & 217                                          & Technology Total:                               & 191                                          \\ \hline
\end{longtable}
\normalsize 

\begin{table}[h]
\centering
\scriptsize
\centering
\caption{Handcrafted Stop Word List. Note that these words are inspected and excluded by manually visiting the first 30 words in the word topics inferred by the model.}
\begin{tabular}{|l|l|l|l|l|l|l|}
\hline
day       & inside   & meaning & reveals & things   & winning & minute \\ \hline
dos       & issue    & meet    & romney  & tips     & frances & colors \\ \hline
double    & jennifer & minutes & ryan    & today    & small   & grows  \\ \hline
easy      & joe      & month   & share   & top      & japan   & green  \\ \hline
essence   & johns    & nation  & shows   & trick    & china   & simple \\ \hline
exclusive & julie    & needed  & special & ultimate & stuff   & autumn \\ \hline
eye       & kate     & obama   & steve   & undos    & pages   & design \\ \hline
faces     & klein    & ons     & stop    & ups      & mitt    &        \\ \hline
falling   & losing   & picks   & stories & wanted   & week    &        \\ \hline
free      & lost     & preview & summer  & ways     & i       &        \\ \hline
good      & makeover & rated   & takes   & white    & super   &        \\ \hline
great     & making   & real    & talking & work     & perfect &        \\ \hline
guide     & matter   & red     & tells   & year     & spring  &        \\ \hline
\end{tabular}
\end{table}
\normalsize \label{table:stop_words}

\bibliographystyle{ACM-Reference-Format-Journals}
\bibliography{acmsmall-my-report}

\end{document}